\documentclass[12pt]{article}
\pdfoutput=1
\usepackage{jheppub, amsmath,amssymb}

\usepackage{color,colortbl}
\usepackage[american]{babel}
\usepackage[utf8]{inputenc}
\usepackage{microtype}
\usepackage{graphicx}
\usepackage{physics}
\usepackage{bbm}
\usepackage{array}
\usepackage{tikz}
\usepackage{amscd}
\newcommand{\cK}{\mathcal{K}}

\newcommand{\cM}{\mathcal{M}}
\newcommand{\cN}{\mathcal{N}}

\newcommand{\cW}{\mathcal{W}}

\newcommand{\bI}{\mathbb{I}}

\newcommand{\bZ}{\mathbb{Z}}

\newcommand{\be}{\begin{equation}}
\newcommand{\ee}{\end{equation}}
\newcommand{\bea}{\begin{equation}\begin{aligned}}
\newcommand{\eea}{\end{aligned}\end{equation}}
\newcommand{\wt}{\widetilde}

\DeclareMathOperator{\Pf}{Pf}

\newcommand{\ba}{\begin{array}}
\newcommand{\ea}{\end{array}}
\newcommand{\bpic}{\begin{tikzpicture}}
\newcommand{\epic}{\end{tikzpicture}}

\renewcommand\Re{{\mathrm{Re}}}

\newcommand{\M}{{\mathfrak M}}

\newcommand{\Llra}{\Longleftrightarrow}
\newcommand{\ra}{\rightarrow}

\newcommand{\lra}{\leftrightarrow}

\newcommand\Xt{\tilde X}

\renewcommand{\k}{\kappa}

\title{Mildly Flavoring domain walls in SU(N) SQCD: baryons and monopole superpotentials}

\author[1,2]{Sergio Benvenuti}
\author[1]{Paolo Spezzati}

\affiliation[1]{International School of Advanced Studies (SISSA), Via Bonomea 265, 34136 Trieste, Italy}
\affiliation[2]{INFN, Sezione di Trieste, Via Valerio 2, 34127 Trieste, Italy}
\emailAdd{benve79@gmail.com, pspezzat@sissa.it}

\abstract{We study supersymmetric domain walls of four dimensional $SU(N)$ SQCD with $N$ and $N+1$ flavors. In $4d$ we analyze the BPS differential equations numerically. In $3d$ we propose the $\cN=1$ Chern-Simons-Matter gauge theories living on the walls.

Compared with the previously studied regime of $F<N$ flavors, we encounter a couple of novelties: with $N$ flavors, there are solutions/vacua breaking the $U(1)$ baryonic symmetry; with $N+1$ flavors, our $3d$ proposal includes a linear monopole operator in the superpotential.}

\begin{document}

\maketitle

%\tableofcontents

%%%%%%%%%%%%%%%%%%%%%%%%%%%%%%%%%%%%%%%%%%%%%%%%%%%%%%%%%%%%%%%%%%%%%%%%%%%%%%%%%%%%%%%%%%%%%%%%%%%%%%%%%%%%%%%%%%%%%%%%%%%%%%%%%%%%%%%%%%%%%%%%%%%%%%%%%%%%%%%%%%%%%%%%%%%%%%%%%%%%%%%%%%%%%%%%%%%%%%%%%%%%%%%%%%%%%%%%%%%%%%%%%%%%%%%%%%%%%%%%%%%%%%%%

\section{Introduction and summary}
The study of space-time defects in Quantum Field Theories has been teaching many important lessons. In this paper, we focus on objects localized on a codimension-$1$ subspace, namely supersymmetric domain walls of $4$-dimensional supersymmetric QCD.
 
Domain walls of $4d$ supersymmetric QFT's have  been studied in \cite{Chibisov:1997rc, Dvali:1996xe, Kovner:1997ca, Witten:1997ep, Smilga:1997cx, Kogan:1997dt, Smilga:1998vs, Kaplunovsky:1998vt, Dvali:1999pk, deCarlos:1999xk, Gorsky:2000ej, Binosi:2000jb, deCarlos:2000jj, Acharya:2001dz, Smilga:2001yz, Ritz:2002fm, Ritz:2004mp,Armoni:2009vv,Dierigl:2014xta,Draper:2018mpj,Hsin:2018vcg,Bashmakov:2018ghn,Delmastro:2020dkz,Benvenuti:2021yqv}. Acharya and Vafa \cite{Acharya:2001dz} proposed the $3d$ Topological Quantum Field Theory (TQFT) living on the $k$-wall of pure $SU(N)$ SYM. \cite{Bashmakov:2018ghn} considered $SU(N)$ and $Sp(N)$ SQCD in the regime with number of flavors $F$ smaller than the dual Coxeter number $h$, $F < h$. 

 In \cite{Benvenuti:2021yqv} we covered the case of $Sp(N)$ with number of flavors $F=h, h+1$. This paper, a companion of \cite{Benvenuti:2021yqv}, does the same for $SU(N)$ gauge group, adding one or two flavors to \cite{Bashmakov:2018ghn}.

Our strategy, as in \cite{Bashmakov:2018ghn, Benvenuti:2021yqv}, is split into two parts: in $4d$, we study the differential equations numerically defining the BPS wall, in $3d$ we propose an $\cN=1$ Supersymmetric Conformal Field Theory describing the $k$-walls.

At low energies, massless $4d$ $SU(N)$ SQCD with $N$ or $N+1$ flavors confines and is described by a Wess-Zumino model. Upon turning on a mass deformation at small masses, we use the Wess-Zumino model to classify the solutions of the differential equations. For $k < \frac{N}{2}$, $k+1$ such solutions exists, each solution hosts a trivial TQFT and a non-trivial Non Linear Sigma Model. At large $4d$ masses instead, the theory reduces to pure $SU(N)$ SYM, so there is only one vacuum for each $k$, hosting the non-trivial TQFT of \cite{Acharya:2001dz}, $U(k)^{\cN=1}_{N-\frac{k}{2},N}$.\footnote{The $k$-walls with $\frac{N}{2}< k < N$ are obtained from the $k$-walls with $1 \leq k\leq\frac{N}{2}$ by $4d$ parity-reversal.}

In $3d$, our task is to find a minimally supersymmetric, $\cN=1$, gauge theory with a mass deformation proportional to $m_{3d}$, such that at $m_{3d}>0$ there is only one supersymmetric vacuum hosting the $U(k)^{\cN=1}_{N-\frac{k}{2},N}$ TQFT, while at $m_{3d}<0$ there are $k+1$ supersymmetric vacua with the correct NLSM's. At zero $3d$ mass ($m_{3d}=0$) there is a $3d$ SCFT, describing the transition of $k+1$ vacua into a single vacuum. Recent progress in the study of $3d$ $\cN=1$ gauge theories includes  \cite{Bashmakov:2018wts, Benini:2018umh, Gaiotto:2018yjh, Benini:2018bhk, Choi:2018ohn, Bashmakov:2018ghn, Benvenuti:2019ujm, Aharony:2019mbc, Sharon:2020xod, Benvenuti:2021yqv}.  Since the $k$-wall is also the parity reversed $(h-k)$-wall, in $3d$ there is a non-trivial duality between the SCFT describing the $k$-wall and the parity reversed SCFT describing the $(h-k)$-wall. 

A typical tool used in the $3d$ analysis is the study of the vacua of the mass-deformed theory, which are matched across the dual description of the IR SCFT. In this paper, we push this tool to its current limits: our proposed $3d$ SCFT's are indeed \emph{strongly coupled}, and the vacua do not exactly match across the dualities. More precisely, as first seen in \cite{Benvenuti:2021yqv} for the BPS walls of $4d$ $Sp(N)$ with $N+2$ flavors, on one side of the duality there are additional vacua if $m_{3d}<0$. This mismatch may be due to the presence of quantum phases similar to the non-supersymmetric quantum phases of QCD domain walls \cite{Komargodski:2017keh, Gaiotto:2017tne}. %We leave a further study of the behavior of  \emph{strongly coupled} $3d$ $\cN=1$ gauge theories to future work.

\vspace{0.2cm}

Let us end this introduction by describing our results in a bit more detail.

\subsubsection*{Domain walls of $4d$ $SU(N)$ with $N$ flavors}
In Section \ref{secF=N}, we analyze the BPS equations in the quantum deformed moduli space description of the theory. The differential equations for the $k$-wall ($k<\frac{N}{2}$) lead to a single \emph{non-baryonic} solution (already found in \cite{Ritz:2004mp,Ritz:2002fm}) and $k$ \emph{baryonic} solutions. In the non-baryonic solution, only the mesons have a non-zero profile along the domain wall, while in the baryonic solutions, both the mesons and the baryons have a non-zero profile. Such baryonic solutions are qualitatively new and have vanishing Witten-Index.

In $3d$ our proposal for the gauge theory on the $k$-wall is $U(k)_{\frac{N-k}{2},\frac N2}^{\cN=1}$ with $N$ fundamentals and a quartic superpotential. Such proposal is a direct generalization of the ones in \cite{Bashmakov:2018ghn}. One difference with respect to \cite{Bashmakov:2018ghn} is the presence of "baryonic vacua", breaking the topological (or magnetic) $U(1)$ symmetry and hosting the NLSM $S^1 \times Gr(J,F)$, where $J=0,\ldots,k-1$ and $Gr(J,F)$ is the Complex Grassmannian. Another difference with respect to \cite{Bashmakov:2018ghn} is that for $k>\frac{N}{2}$, the semiclassical analysis of the mass deformed theory yields additional vacua. This is the same behaviour of $Sp(N)$ with $F=N+2$ flavors in \cite{Benvenuti:2021yqv} and we expect that for $k>\frac{N}{2}$ the $3d$ quantum filed theory is \emph{strongly coupled} and such additional vacua are not related to the phase transition relevant for the domain wall.\footnote{The value $k=\frac N2$, for $N$ even, is special: we numerically find only two domain wall solutions instead of the expected $\frac N2+1$ vacua. A similar situation has been studied by \cite{Benvenuti:2021yqv} when considering $Sp(N)$ with $F=N+2$ flavors domain walls, that have the same property of connecting vacua that lie on the real axis. In that situation, the seemingly single solution, was argued to be a superposition of different solutions. We believe that this is also the case for $k=\frac N2$ domain walls of $SU(N)$ with $N$ flavors. Moreover, if $k=\frac N2$, the $3d$ analysis is reliable and produces $\frac{N}2+1$ vacua, further supporting such expectation.} 

\subsubsection*{Domain walls of $SU(N)$ with $N+1$ flavors}
In Section \ref{secF=N+1}, we analyze the mass deformed S-confining description of the theory and find $k+1$ solutions in $4d$, with global symmetry $U(F)$ is broken to $U(J) \times U(F-J)$, $J=0,\ldots,k$. All solutions are non-baryonic (we provide an argument for the absence of baryonic solutions in this case) and host the corresponding non-trivial NLSM with a Complex Grassmannian target space.

In $3d$ our proposal is $U(k)^{\cN=1}_{(N-k)/2, N/2}$ with $N+2$ fundamentals, and global symmetry $SU(N+1) \times U(1)$. The structure of this proposal is qualitatively different from \cite{Bashmakov:2018ghn},\footnote{A direct generalization of the proposals in \cite{Bashmakov:2018ghn} does not satisfy the required infrared dualities. Moreover, as we detail in Appendix \ref{sec:DG}, the vacua of the mass deformed theory would host a non-trivial TQFT, but since the $4d$ description is in terms of a Wess-Zumino model, there should be no non-trivial TQFT factors.} since there is a monopole operator in the superpotential, breaking the topological $U(1)$ symmetry explicitly. Moreover, there are $F+1$ fundamental fields, instead of $F$ as for the cases of $SU(N)$ with $F\leq N$ flavors. Encouraging evidence for our proposal with the monopole superpotential comes from $\cN=2$ dualities \cite{Nii:2020xgd, Benini:2017dud}, which can be deformed to the $\cN=1$ dualities that should be satisfied by the $3d$ theory living on the domain walls.

The full superpotential of the $3d$ theory includes many quartic terms compatible with the global symmetry. Accordingly, the analysis of the vacua in the mass deformed theory is involved; the answer depends on which region of the parameters space we are in. We show that at negative $m_{3d}$ the expected vacua are present (even if additional vacua also appear), while at positive $m_{3d}$ there is only the expected vacuum with $U(k)^{\cN=1}_{N-k/2, N/2}$ TQFT. A more thorough analysis of the vacua of the $3d$ mass deformed theories (which probably requires also an understating of strong coupling phenomena) is beyond the scope of this paper.

 %%%%%%%%%%%%%%%%%%%%%%%%%%%%%%%%%%%%%%%%%%%%%%%%%%%%%%%%%%%%%%%%%%%%%%%%%%%%%%%%%%%%%%%%%%%%%%%%%%%%%%%%%%%%%%%%%%%%%%%%%%%%%%%%%%%%%%%%%%%%%%%%%%%%%%%%%%%%%%%%%%%%%%%%%%%%%%%%%%%%%%%%%%%%%%%%%%%%%%%%%%%%%%%%%%%%%%%%%%%%%%%%%%%%%%%%%%%%

\section{BPS domain walls of $SU(N)$ with $N$ flavors}
\label{secF=N}
In this section we consider $\mathcal{N}=1$ SQCD with gauge group $SU(N)$ and $N$ flavors of quarks $Q_I$, $\wt Q^I$ in the (anti)-fundamental representation. The UV massless model has zero superpotential.

Let us review the well known IR behavior of the model \cite{Taylor:1982bp, Seiberg:1994bz, Seiberg:1994pq}.  The non-anomalous continuous global symmetry group is $SU(F)\times SU(F)\times U(1)_B\times U(1)_R$. The gauge-invariant operators which describe the moduli space of the massless theory (at the classical level) are the mesons $M_I^J=Q_I^{\alpha}\tilde{Q}^J_{\alpha}$, the baryon $B=\epsilon^{I_1\dots I_{N}}\epsilon_{\alpha_{1}\dots\alpha_{N}}Q^{\alpha_1}_{I_1}\cdots Q^{\alpha_{N}}_{I_{N}}$ and anti-baryon $\tilde{B}=\epsilon_{I_1\dots I_{N}}\epsilon^{\alpha_{1}\dots\alpha_{N}}\tilde{Q}_{\alpha_1}^{I_1}\cdots \tilde{Q}_{\alpha_{N}}^{I_{N}}$.    
 The massless theory has classical moduli space which is singular at the origin when $M_I^J=B=\wt B=0$. However at the quantum mechanical level there is a constraint 
\begin{equation}
\label{eq:contraintSUN}
\det M- B\tilde{B}=\Lambda^{2N}.
\end{equation}
in terms of the dynamically generated scale $\Lambda$. This constraint smooths out the singularity at the origin, yielding a smooth moduli space. This smooth manifold will be called $\cM_{N,N}$.\footnote{
In the special case of $SU(2)$, for which the fundamental representation coincides with the anti-fundamental, this is precisely the constraint $\Pf M_{Sp(1)}=\Lambda^{2N}$ which appears in the $Sp(1)$ model with $2$ flavors. This can be seen reorganizing the baryons and the mesons into the matrix 
\begin{equation}
 M_{Sp(1)}=\begin{pmatrix}
0& B&M_{11}&M_{12}\\
-B&0&M_{21}&M_{22}\\
-M_{11}&-M_{21}&0&\tilde{B}\\
-M_{12}&-M_{22}&-\tilde{B}&0
\end{pmatrix}
\end{equation}
that maps the $Sp(1)$ mesons to the $SU(2)$ mesons and baryons.}
 
We then turn on a diagonal mass term for the quarks,
\be 
W_m=m_{4d}\Tr M.
\ee 
This breaks the flavor symmetry $SU(F)\times SU(F)\rightarrow SU(F)$ and it leaves only a discrete R-symmetry unbroken, i.e. $\bZ_{2N}$. We can implement the quantum constraint on the would-be moduli space using a Lagrange multiplier $A$. The low-energy physics is described by the effective superpotential
\begin{equation}
\label{eq:SUNpotential}
W=m_{4d}\Tr M+ A(\det M- B \tilde{B}-\Lambda^{2N}),
\end{equation}

The solutions of the F-term equations are $N$ gapped vacua with R-symmetry breaking $\bZ_{2N}\rightarrow \bZ_2$:
\begin{equation}
M=\tilde{M}\bI_N, \quad \tilde{M}^N=\Lambda^{2N}, \quad B=\tilde{B}=0,\quad A=\frac{m_{4d} \tilde{M}}{\Lambda^{2N}}.
\end{equation}
We can see that the moduli space is lifted when $m_{4d} \neq 0$. 
When the quark mass is large $\abs{m_{4d}}\gg\Lambda$, one can integrate out the quarks and remain with pure SYM, which has, in turn, $N$ vacua. On the other hand, if $\abs{m_{4d}}\ll\Lambda$ the effective description as a WZ model on the moduli space is reliable.

In the following we will not use the formulation with the Lagrange multiplier field $A$, we will instead  study the NLSM with target space $\cM_{N,N}$, embedded in the flat $(N^2+2)$-dimensional complex space parameterized by $M, B, \tilde{B}$. As in the $Sp(N)$ $F=N+1$  case \cite{Benvenuti:2021yqv}, we do not really need complete information about the Kähler potential, except for the fact that it is smooth, it is natural to use the canonical, quadratic, Kähler potential for the ambient space
\be\label{cankal} \cK= M_I^J (M^{\ast})_J^I + B B^{\ast} + \tilde{B}\tilde{B}^{\ast}\,.\ee

\subsection{Analysis of the BPS equations}\label{BPS1}
In this paper we study domain walls of $\cN=1$ $4d$ massive SQCD that preserve half of the supersymmetry, these are called BPS domain walls \cite{Dvali:1996xe,Abraham:1990nz,Cecotti:1992rm}. These walls interpolate between two vacuum configuration at the two ends of the Universe, at $x=\pm\infty$.  When the low energy physics of the model is described by a Wess-Zumino model with superpotential $W$, the tension of the walls is fixed by supersymmetry and it is given by
\be
T=2\abs{W(v_i)-W(v_j)},
\ee
the difference of the superpotential being evaluated at the two vacua $v_i$, $v_j$ at the two ends of the Universe. Moreover, if we are considering a Wess-Zumino model there are first order differential equations to study the trajectory of such domain walls \cite{Fendley:1990zj,Abraham:1990nz}:
\begin{equation}
\label{eq:diff}
\partial_{x}\Phi^a=e^{i\gamma}\mathcal{K}^{a\bar{b}}\overline{\partial_b W},
\end{equation}
where $\Phi^a$ are the chirals of the WZ model, $\mathcal{K}^{a\bar{b}}$ is the inverse Kähler metric and $e^{i\gamma}=\frac{\Delta W}{\abs{\Delta W}}$. The trajectory of the domain wall in the W-space, that is the image of $W(\Phi^a)$ along the domain wall solution, is a straight line 
\begin{equation}
\label{eq:Wline}
\partial_{x}W=e^{i\gamma}\abs{\partial W}^2.
\end{equation}
Let us point out that the very existence of the domain walls does not depend on the D-terms \cite{Cecotti:1992rm}. In other words, it is insensitive to the choice of the Kähler metric. This will allow us, in the following, in order to find domain wall solutions, to choose a Kähler metric as we like, provided that the Kähler metric does not have singularities along the domain wall solution.

Let us apply this general formalism to our case of interest, the Wess-Zumino model described at the beginning of this Section. In order to solve the BPS equations \eqref{eq:diff} for a constrained system we need to choose a set of coordinates for the target manifold $\cM_{N,N}$ of the NLSM and rewrite the equations \eqref{eq:diff} in term of these coordinates.\footnote{When choosing the set of coordinates one has to make sure that the domain wall solutions and the vacua of the model can be described in those coordinates. For example if we solve \eqref{eq:contraintSUN} like $B=\frac{ 1}{\tilde{B}} (\text{det}( M) \Lambda^{2N})$ we could not see the vacua because, on the vacua (which have  $\tilde{B}=B=0$), this expression is not valid.}

At this point we use the $SU(F) \times U(1)$ flavor symmetry. The mesonic $SU(F)$ symmetry allows to diagonalize the meson matrix $M=\text{diag}(\xi_1,\dots,\xi_N)$. Since  the equations \eqref{eq:diff} transform covariantly under the flavor symmetry, if at some point along the domain wall solution we set the off-diagonal components of $M$ to zero, they remain zero along all the solution. Similarly, using the baryonic $U(1)$ flavor symmetry we set  $\tilde{B}=B$. In this way we reduce the number of independent functions from $N^2+2$ to $N+1$.

We thus study a reduced system with variables $\xi_i$ and $B$ (we express $\xi_i$ in units of $\Lambda^2$, $B$ in units of $\Lambda^N$), satisfying the constraint 
\be \label{eq:contraintSUN1} \prod_{i=1}^N \xi_i = B^{2} +1\,, \ee
providing the embedding into a $N+1$ dimensional target space with flat metric, i.e. with Kähler potential 
\be \cK=\sum_{i=1}^N  | \xi_i |^2 +2 |B|^2\,.\ee

The superpotential on the reduced target space is just given by the mass term
\be \cW= \sum_{i=1}^N \xi_i \ee

% and setting $A=\frac{m_{4d}\Lambda^2 }{\Lambda^{2N}}$ to one,  the superpotential becomes
%\be\label{eq:superpotential barion} W=\sum_i\xi_i+\bigl(\prod_i \xi_i-B\tilde{B}-1\bigr).
%\ee

%The differential equations we get with the ansatz that the meson matrix $M$ has diagonal form are the same equations we can derive by considering our system to be given by the eigenvalues $\xi_i$ and the superpotential \eqref{eq:superpotential barion}. 

%We call this system the associated reduced system and we  will study it. The Kähler potential \eqref{cankal} takes the form in terms of the  fields $\xi_i$, $B$ and $\tilde{B}$, 

%It is much easier to solve these differential equations if we make some further assumptions that simplify the problem.  First of all, we make the ansatz  $\tilde{B}=B$. This ansatz is coherent with the equations \eqref{eq:diff}, which is not a surprise because the theory enjoys the $U(1)$ baryonic symmetry. 

The next step to take is to solve the constraint \eqref{eq:contraintSUN1}. We write 
\be
\label{eq:constlambda1}
\xi_1=\frac{1+B^2}{\prod_{i\neq1}\xi_i}.
\ee
This particular choice of solving the constraint allow us to have well defined coordinates on the vacua of the theory, where the baryons vanish. 
%The superpotential in these coordinates reads
%\be
%W=\sum_{i=2}^N\xi_i+\frac{1+B^2}{\prod_{i\neq1}\xi_i}.
%\ee
Then, projecting the canonical Kähler potential and flat metric of the ambient space $\mathbb{C}^{N+1}$ on the manifold determined by the constraint, we get  the induced Kähler potential 
\be
\cK=\sum_{i=2}^N\abs{\xi_i}^2+\abs{B }^2+\frac{\abs{1+B^2}^2}{\prod_{i\neq1}\abs{\xi_i}^2},
\ee
whereas the metric reads
\begin{multline}\label{Kinduced}
\cK_{a\bar{b}}=\sum_{i=2}^N\abs{d\xi_i}^2\biggl(1+\frac{\abs{1+B^2}^2}{\prod_{j\neq1}\abs{\xi_j}^2\abs{\xi_i}^2}\biggr)+\sum_{i\neq j}d\xi_i^{\ast}d\xi_j\frac{\abs{1+B^2}^2}{\prod_{h\neq1}\abs{\xi_h}^2\xi_j\xi_i^{\ast}}+\\
+2\sum_{i=2}^N\frac{B^{\ast}(1+B^2)dB^{\ast}d\xi_i}{\prod_{h\neq1}\abs{\xi_h}^2\xi_i}+2\sum_{i=2}^N\frac{B(1+B^{\ast 2})dBd\xi_i^{\ast}}{\prod_{h\neq1}\abs{\xi_h}^2\xi^{\ast}_i}+\abs{dB}^2\biggl(1+\frac{4\abs{B}^2}{\prod_{h\neq1}\abs{\xi_h}^2}\biggr).
\end{multline}

The final expression for the differential equations \eqref{eq:diff} is involved and we will not write it down here, however, all the elements to write an efficient code to solve the differential equation have been given. What we have just described is the general procedure to solve the differential \eqref{eq:diff} in case of a WZ model with non-trivial target manifold for the chiral fields. %However, in case tha target manifold is one dimensional we can avoid to solve the differential equations \eqref{eq:diff}, finding the  domain wall solution algebraically. We will describe this procedure in the next section. 
We now turn to describing the solutions we have found.

\subsection*{Non-baryonic walls}
To ease our way into the study of $SU(N)$ with $N$ flavors domain walls, we start with the case in which the baryons are spectators, that is $B=\tilde{B}=0$ along the whole solution. This type of solutions were already found in \cite{Ritz:2004mp,Ritz:2002fm}. 

%Using the $SU(F)$ flavor symmetry we can diagonalize the meson matrix $M$. Notice that if $M(x)$ is diagonal at a certain point $x_0$, the form of the differential equations   \eqref{eq:diff} imply that $M(x)$ is diagonal for any $x$, so we only need to  study $M$'s diagonal along the whole solution.

\begin{figure}[h!]
\centering
 \includegraphics[width=.32\textwidth]{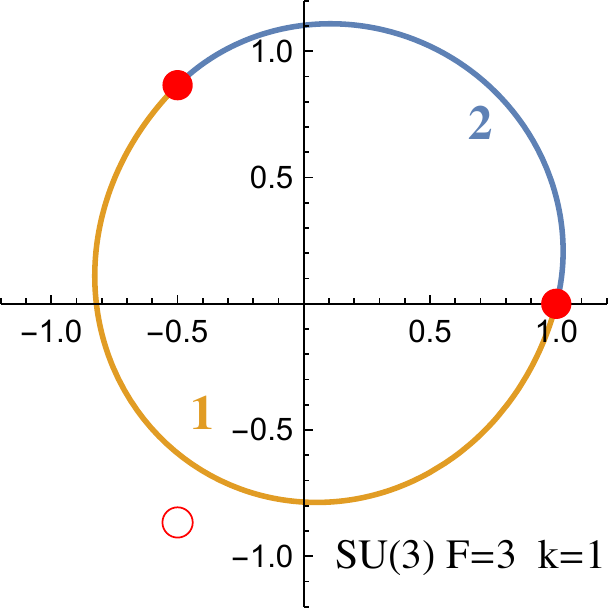}\hspace{\stretch{1}}
  \includegraphics[width=.32\textwidth]{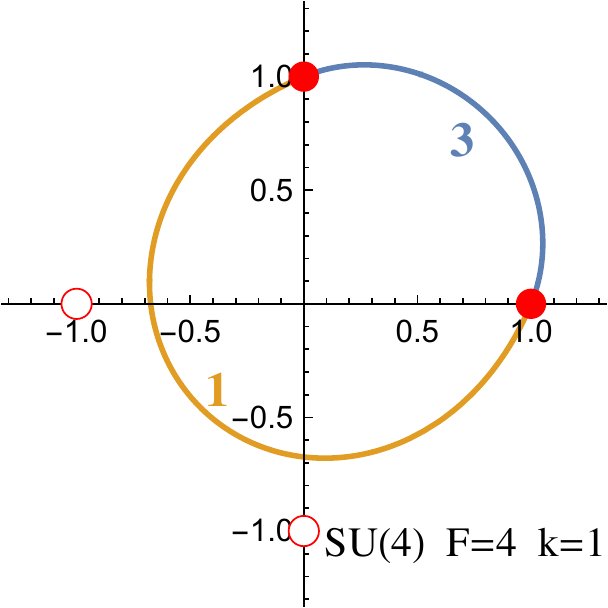}\hspace{\stretch{1}}
    \includegraphics[width=.32\textwidth]{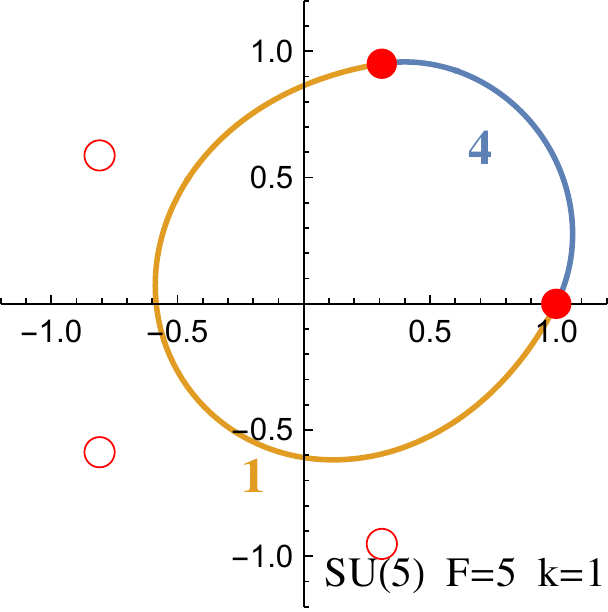}\hspace{\stretch{1}}
 \caption{Examples of 1-wall solutions for $SU(N)$ with $F=N$ flavors for $N=3,4,5$. In each of these cases the eigenvalues of the meson matrix $M$ are split in two groups of $k_1=1$ and $k_2=N-1$ elements.}
 \label{fig:1wallsu221}
\end{figure}

\begin{figure}[h!]
\centering
\hspace{\stretch{1}}\includegraphics[width=.32\textwidth]{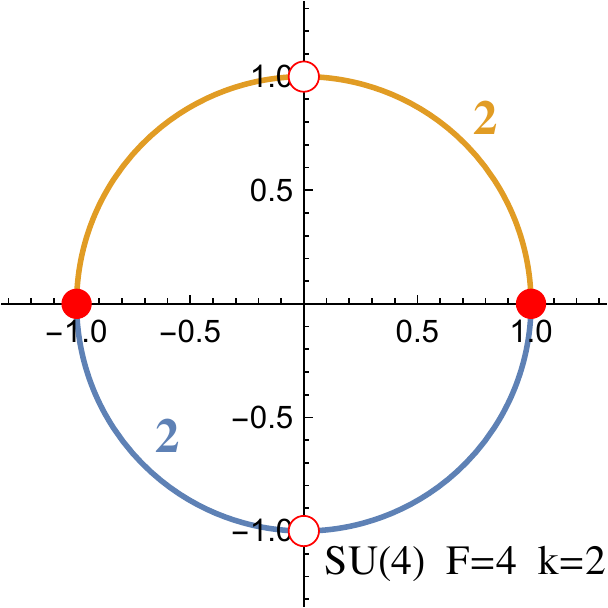}\hspace{\stretch{1}}
\includegraphics[width=.32\textwidth]{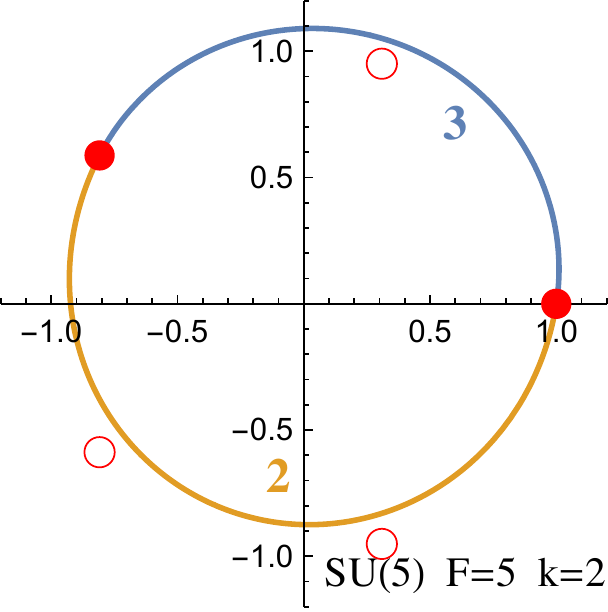}\hspace{\stretch{1}}
 \caption{Examples of 2-wall solution for $SU(N)$ with  $F=N$ flavors, where $N=4,5$.}
 \label{fig:2wallsu443}
\end{figure}

Solving numerically the differential equations \eqref{eq:diff}, we found a single solution, with the property that the $N$ eigenvalues of $M$ are split into two groups: $k$ equal eigenvalues following one path and $N-k$ equal eigenvalues following another path. Therefore, the flavor symmetry is broken along the solutions into  $U(N) \rightarrow U(k)\times U(N-k)$.  Acting with this residual symmetry, one produces a family of solutions parametrized by the Grassmannian $\mathrm{Gr}(k,N)=\frac{U(N)}{U(k)\times U(N-k)}$. Examples of such solutions are sketched in Figure \ref{fig:1wallsu221} and in Figure \ref{fig:2wallsu443}.

It is interesting to note that  solutions where the eigenvalues split into at most two groups can be analyzed more easily algebraically, as we now explain. Let us denote the eigenvalues $\xi_1$ the first group of $k_1$ eigenvalues and $\xi_2$ the second group of $k_2$ eigenvalues. Let us now evaluate the constraint  \eqref{eq:contraintSUN}, hence writing one group of the eigenvalues in terms of the other, obtaining a superpotential that depends only on one superfield $\xi_1$
\be
W=k_1\xi_1+k_2\xi_1^{-\frac{k_1}{k_2}}
\ee
(here the fields $\xi_i$ have been rescaled in units of $\Lambda^2$ and $\Lambda^2m_{4d}$ has been set to one for simplicity).
Now we use the fact the domain wall trajectory in the $W$-space is a straight line, so in order to find the solutions we just have to invert the equation
\be
W\Bigl( M \big|_{x= +\infty} \Bigr) \, t + W \Bigl( M\big|_{x = -\infty} \Bigr) \, (1-t) = k_1 \, \xi_1(t) + k_2 \, \xi_1(t)^{-k_1/k_2}
\ee
 Notice that this algebraic method does not rely on the choice of a Kähler potential. Studying the previous algebraic equation is relatively easy to see that the only solution for the $k$-wall sector has $k_1=k$ and $k_2=N-k$, or viceversa. 

\subsection*{Baryonic walls}
Now we relax the previous assumption that $B=\tilde{B}=0$ along the solution.  We than have to solve the differential equation \eqref{eq:diff}. The general procedure to 'diagonalize' the problem has been described around eq. \eqref{Kinduced}. From the numerical analysis, for each $N$ and $k$, we found $k$ baryonic domain walls solutions, parameterized by $J=0,1,\ldots,k-1$, such that the mesonic eigenvalues split into two groups of $J$ and $k-J$ equal eigenvalues. We did not find any solutions where the eigenvalues split into three or more groups.\footnote{Notice that when the eigenvalues split in two groups of $k_1$ and $k_2$ elements and express each chiral fields as $\xi_i=\rho_i e^{i\phi}$, $B=\beta e^{i\alpha}$. Then to implement the equations \eqref{eq:diff} we need to pay attention: after imposing the constraint \eqref{eq:constlambda1} we can see only $k_1-1$ eigenvalues of the first group. So we need to make sure that also the $k_1$-th element is equal to the other imposing the \eqref{eq:constlambda1} along the differential equations.}

We sum up the solutions found in Table \ref{tab:solution SU N}.

 \begin{table}
 \begin{center}
\begin{tabular}{|c|l|c|}
\hline
Wall & Effective theory& Witten Index\\
\hline
$k$& $\mathrm{Gr}(k,N)$& $\smqty(N\\k)$\\
&$S^1\times \mathrm{Gr}(J,N), \quad J\in \{0,\dots,k-1\}$&$0$\\
\hline
\end{tabular}
\end{center}
 \caption{\label{tab:solution SU N}Domain wall solutions found for 4d $\cN=1$ $SU(N)$ SQCD with $F=N$ flavors in the regime when $m_{4d}\ll\Lambda$. For each $k<\frac N2$-wall sector are included also the various contributions to the Witten index from each solution.} 
\end{table}

\begin{figure}[h!]
\centering
 \includegraphics[width=.32\textwidth]{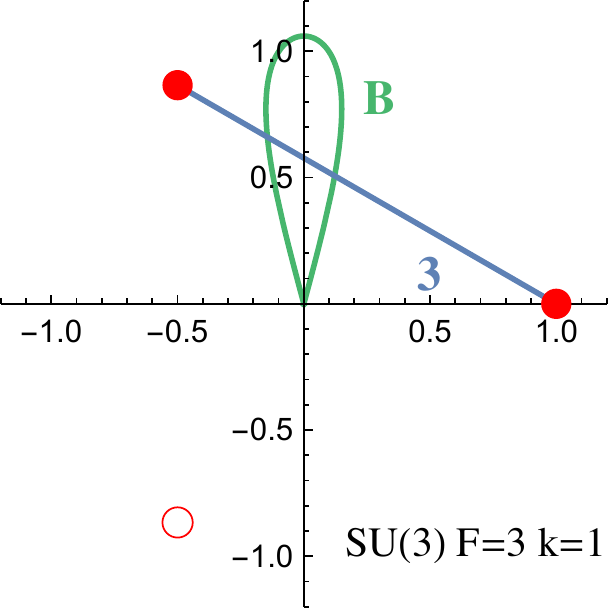}\hspace{\stretch{1}}
  \includegraphics[width=.32\textwidth]{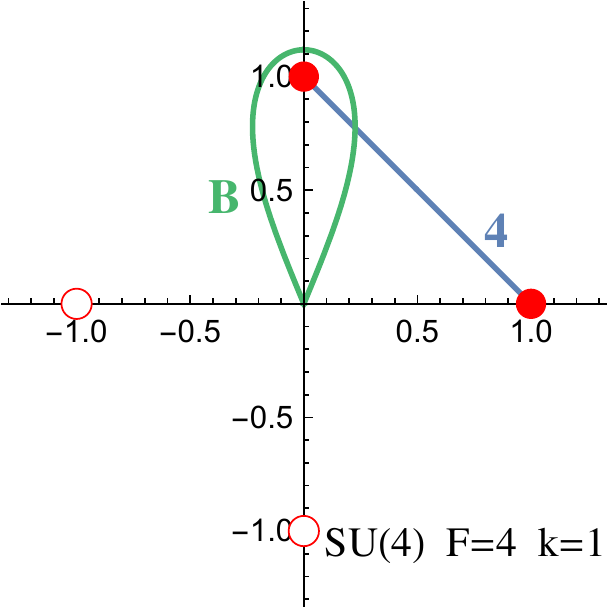}\hspace{\stretch{1}}
   \includegraphics[width=.32\textwidth]{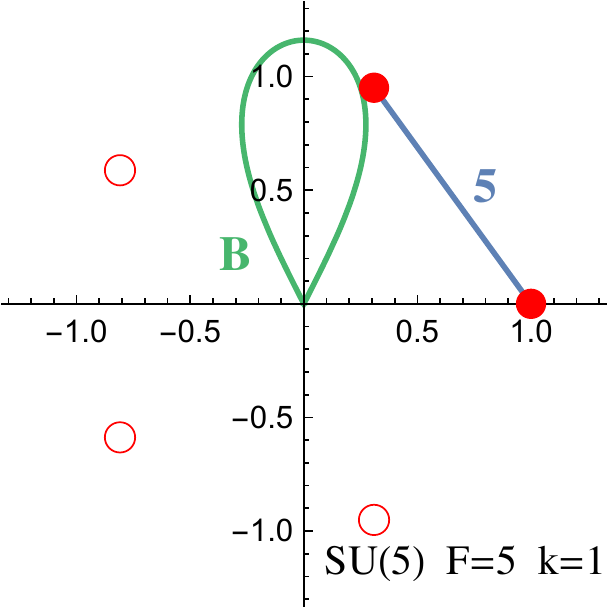}\hspace{\stretch{1}}
 \caption{Examples of domain walls of $SU(N)$ $F=N$: each figure represent a 1-wall solution in which the eigenvalues (in blue) of the meson matrix $M$ do not split. The baryons are depicted in green. The three cases are respectively from left to right  $N=3,4,5$.}
 \label{fig:1wallsu332}
\end{figure}

\begin{figure}[h!]
\centering
\hspace{\stretch{1}} \includegraphics[width=.32\textwidth]{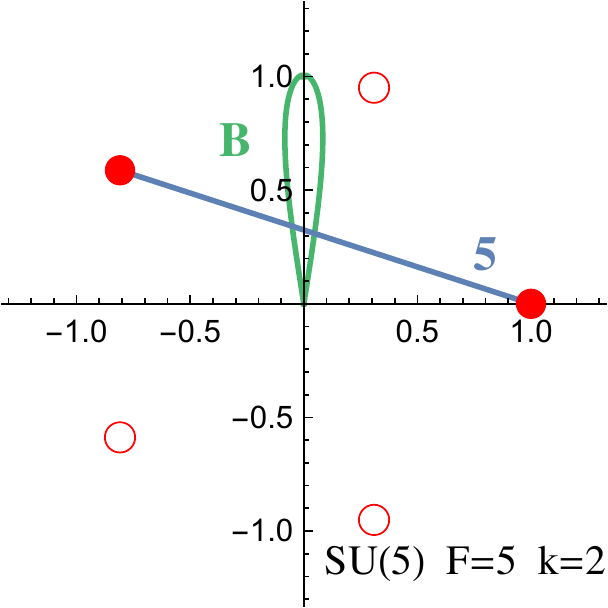}\hspace{\stretch{1}}
 \includegraphics[width=.32\textwidth]{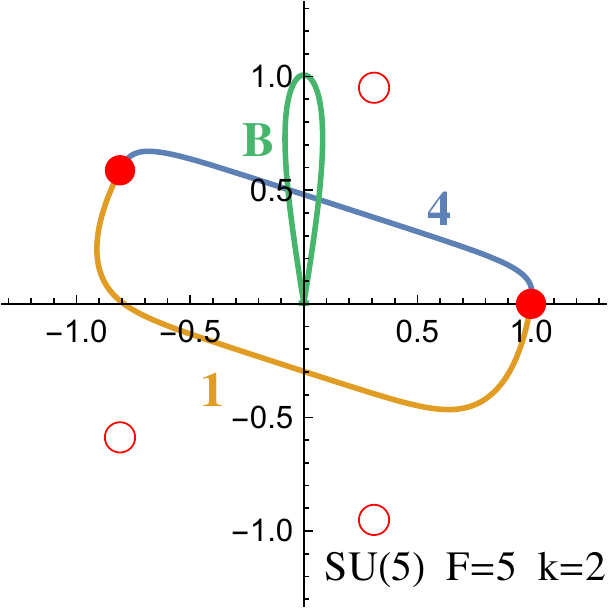}\hspace{\stretch{1}}
 \caption{Examples of  2-wall solutions for $SU(5)$ with $F=5$ flavors.}
 \label{fig:2wallsu441}
\end{figure}

We reproduce various examples of our numerical baryonic solutions in  Figure~\ref{fig:1wallsu332} and Figure~\ref{fig:2wallsu441}. 

Notice that the solutions with $J=0$, as the ones in Figure~\ref{fig:1wallsu332} and the left one in Figure~\ref{fig:2wallsu441}, depend non-trivially only on two functions (the single mesonic eigenvalue and the barion), hence such solutions can also be computed algebraically, using the constraint as for the non-baryonic wall. This provides a reassuring check for the numerical analysis. 

In these solutions the baryonic symmetry is broken because the baryons take VEV, therefore we expect a family of solutions which is parametrized by an $S^1$. This can be traced back to the fact that $B=\tilde{B}$ fixes the difference of the phase. If the flavor symmetry is also broken (as in Figure~\ref{fig:2wallsu441}) the family of solution is described by the Cartesian product of the $S^1$ and  the Grassmaniann $\mathrm{Gr}(k_1,N)$.

\paragraph{Matching the Witten-Index} When $m_{4d} \ll \Lambda$, there are $k+1$ solutions. The $k$ ``baryonic'' domain wall solutions host a NLSM with target space $S^1 \times Gr(J,N)$, hence they have zero Witten index, due to the presence of the $S^1$ factor. The single non-baryonic wall hosts the NLSM with target space $Gr(k,N)$, hence its Witten Index is $\smqty(N\\k)$. This matches the Witten Index of the TQFT $U(k)^{\mathcal{N}=1}_{N-\frac k2, N}$, which describes the $k$-wall when $m_{4d} \gg \Lambda$.

\subsubsection{The parity-invariant walls}
The case $k=\frac N2$ (parity-invariant wall), and $N$ is even, is special because the solutions we have found do not exactly follow the pattern of Table \ref{tab:solution SU N}. In fact, we only found two solutions:  eigenvalues of the meson matrix $M$ were evenly split or not split at all. A similar phenomenon was observed in \cite{Benvenuti:2021yqv} for this kind of domain walls, connecting two vacua where the VEVs of the fields are along the real axis. These domain walls were dubbed ``parity-invariant'' walls. In  \cite{Benvenuti:2021yqv} we showed that there is not a single solution, but there is a superposition of many different solutions. The difference between those solutions can be resolved by deforming the Kähler potential, explicitly but softly breaking the flavor symmetry. We believe that this is the case also in our situation, but we were not able to find such solutions, probably due to the complication of having the baryons involved. This scenario is also supported by the analysis of the vacua of the $3d$ model describing the low energy physics on the wall, where we find $\frac{N}{2}+1$ vacua (this analysis is carried out in the next subsection \ref{SUNeffective}). For $SU(N)$ with $N+1$ flavors we study such ``parity-invariant'' walls in Sec \ref{parityinv}.

\subsection{Effective $3d$ theory on the walls: baryonic vacua}\label{SUNeffective}
We now move to illustrate the 3d theories that can describe the low energy physics trapped on the domain walls found in Section \ref{BPS1}.  We propose as low energy model living on the $k$-wall  the $3d$ $\cN=1$ theory 
\begin{equation}
U(k)_{\frac{N-k}{2},\frac N2}^{\cN=1} \quad \text{with $N$ fundamentals } X, \quad \cW \sim + |X|^4
\label{eq:3dtheory}
\end{equation}
The evidence for this proposal, which is somewhat weaker with respect to the cases of $F < N$ flavors, comes from dualities and from the study of massive vacua. 

The following duality is known for $\cN=2$ theories \cite{Benini:2011mf}:\footnote{This $\cN=2$ duality can be derived directly from Aharony duality for $U(k)$ with $(N,N)$ flavors, giving real masses to $(0,N)$ flavors.} 
\be\label{n=2SUdualityF=N}
 \ba{c} U(k)^{\cN=2}_{\frac{N}{2}} \,\,\,\,\, \text{w/ $(N,0)$ flavors} \\ \cW_{\cN=2}=0 \ea
       \Llra
\ba{c} U(N-k)^{\cN=2}_{-\frac{N}{2}} \,\,\,\,\, \text{w/ $(0,N)$ flavors} \\ \cW_{\cN=2}=\mu \, \M \ea \ee
($\mu$ is a gauge singlet chiral field, $\M$ is a supersymmetric monopole). This duality is well tested, for instance the supersymmetric partition functions are known to match.

In $\cN=1$ language the above duality reads
\be \ba{c} U(k)^{\cN=1}_{\frac{N}{2},\frac N2} \, \text{with adjoint $\Phi$ and $N$ $X$}'s \\ \cW=\Tr(X^i \Phi X^\dagger_i) -\frac{N}{4} \Tr(\Phi^2) \ea
           \Llra
\ba{c} U(N-k)^{\cN=1}_{-\frac{N}{2},-\frac N2} \, \text{with adjoint $\Psi$ and $N$ $\Xt$}'s \\ 
\cW=\Tr(\Xt^i \Psi \Xt^*_i) + \frac{N}{4} \Tr(\Psi^2) + \\ + (\mu \, \M+ c.c.) \ea \ee
Integrating out the massive adjoints $\Phi$ and $\Psi$ we can write the $\cN=2$ duality as
\be \label{N=2rewitten}\ba{c} U(k)^{\cN=1}_{\frac{N-k}{2},\frac N2} \, \text{with $N$ $X$}'s \\ 
\cW=\frac{1}{2N} \Tr(X^i X^\dagger_j X^j X^\dagger_i) \ea
           \Llra
\ba{c} U(N-k)^{\cN=1}_{-\frac{k}{2},-\frac N2} \, \text{with $N$ $\Xt$}'s \\
 \cW=-\frac{1}{2N}  \Tr(\Xt^i \Xt^\dagger_j \Xt^j \Xt^\dagger_i) + (\mu\, \M + c.c.)\ea \ee
 
Let us enphasize that \ref{N=2rewitten} is just \ref{n=2SUdualityF=N} written in a different notation. We can now deform the $\cN=2$ duality to an $\cN=1$ duality, turning on 
 \be \delta \cW \sim \Tr(X^i X^\dagger_i)^2 \Llra  \delta \cW \sim\Tr(\Xt^i \Xt^\dagger_i)^2 + \mu \mu^* \ee
 
Along with the $\cN=1$ superpotential deformation quartic in the fundamental, also the deformation $\delta \cW = \mu \mu^*$ is generated (since it does not violate any global symmetry and with only $\cN=1$ supersymmetry there are no non-renormalization theorems), so the gauge singlet field $\mu$ in \ref{n=2SUdualityF=N} becomes massive.

We get, in analogy to the case $F<N$, an $\cN=1$ duality enjoyed by \ref{eq:3dtheory}:
\be\label{n=1SUdualityF=N}
 \ba{c} U(k)^{\cN=1}_{\frac{N-k}{2}, \frac{N}{2}} \,\,\,\,\, \text{with $N$ fundamentals } X, \\
  \cW \!= \!   \Tr(X^i X^\dagger_j X^j X^\dagger_i) +  \Tr(X^i X^\dagger_i)^2 \ea
     \!\!\!  \Llra \!\!\!
\ba{c} U(N-k)^{\cN=1}_{-\frac{k}{2}, -\frac{N}{2}} \,\,\,\,\, \text{with $N$ fundamentals } \tilde{X}, \\ 
\cW \! = \! -\Tr(\Xt^i \Xt^\dagger_i)^2-  \Tr(\Xt^i \Xt^\dagger_j \Xt^j \Xt^\dagger_i) \ea \ee

The duality \ref{n=1SUdualityF=N} is precisely what we expect from the $4d$ equivalence between a $k$-wall and a time reversed $(N-k)$-wall.

Another piece of evidence comes from the $\cN=2$ $SU \lra U$ duality \cite{Aharony:2014uya}
\be \label{AF1} \ba{c} SU(N)^{\cN=2}_{-1-N/2} \,\,\,\,\, \text{w/ $(N,0)$ flavors} \\ \cW_{\cN=2}=0 \ea
\Llra
\ba{c} U(1)^{\cN=2}_{N/2} \,\,\,\,\, \text{w/ $(0,N)$ flavors} \\ \cW_{\cN=2}=0 \ea \ee
The $\cN=1$ deformation of the above duality is expected to be
\be \label{SUU1} \ba{c} SU(N)^{\cN=1}_{-1} \,\,\,\,\, \text{w/ $N$ fundamentals $Y$} \\ \cW \sim - Y^4 \ea
\Llra
\ba{c} U(1)^{\cN=1}_{N/2} \,\,\,\,\, \text{w/ $N$ fundamentals $X$} \\ \cW \sim + X^4 \ea \ee
This duality describes the equivalence between the interface theory $SU(N)^{\cN=1}_{-1}$ and the 1-wall theory $U(1)_{N/2}$.\footnote{ This extends the $1$-wall $\lra$ interface duality of \cite{Bashmakov:2018ghn}
\be \label{SUU1f} \ba{c} SU(N)^{\cN=1}_{-1+F/2-N/2} \,\,\,\,\, \text{w/ $F$ fundamentals}  \ea
\Llra
\ba{c} U(1)^{\cN=1}_{N-F/2} \,\,\,\,\, \text{w/ $F$ chirals}  \ea \ee
 to the case of $F=N$.}

So far the story seems equivalent to the cases of $4d$ $SU(N)$ with $F<N$ flavors of \cite{Bashmakov:2018ghn} and $Sp(N)$ with $F \leq N+1$ flavors of \cite{Bashmakov:2018ghn, Benvenuti:2021yqv} . The difference is that for $4d$ $SU(N)$ with $N$ flavors, when we study the massive vacua of the $3d$ $\cN=1$ models, we do not find perfect matchings between dual $3d$ theories in \ref{n=1SUdualityF=N} and \ref{SUU1}, and with the $4d$ analysis of the previous section. More precisely, the vacuum structure of \eqref{eq:3dtheory} does match the $4d$ analysis if $k < \frac N2$, but it does not  if $k>\frac N2$ (the $3d$ theory \eqref{eq:3dtheory} has additional vacua not seen in $4d$ or in the $3d$ dual). As for $Sp(N)$ with $F = N+2$ in \cite{Benvenuti:2021yqv}, this phenomenon should be due to strong coupling effects present in our $3d$ models when $k>\frac N2$.

\subsubsection*{Analysis of the massive vacua}
Let us now study semi-classically the vacuum structure of \eqref{eq:3dtheory}. The full superpotential generated by the RG flow has the form
\begin{equation}
\label{eq:superpotential}
\mathcal{W}=\frac{1}{4}\Tr(XX^{\dag}XX^{\dag})+\frac{\alpha}{4}\Tr(XX^{\dag})^2+m \Tr XX^{\dag}.
\end{equation}
We assume that our SCFT lies in  a region of the parameter space with $\alpha>-\frac1 k$. Since the precise value of $\alpha$ does not change the results, hence-forth we set  $\alpha=0$ for simplicity.

The analysis of the vacua is carried out more or less in the same fashion as the $Sp(N)$ cases in \cite{Benvenuti:2021yqv}. We diagonalize the matrix $XX^{\dag}=\text{diag}(\lambda_1^2,\dots,\lambda_N^2)$ using the flavor and gauge symmetry. Since $XX^{\dag}$ is semi-positive definite, $\lambda_i^2\ge0$. Supersymmetric vacua satisfy the F-term equations
\be
\lambda_i(\lambda_i^2+m)=0\quad  i\in \{0,\dots,k\}.
\ee
These equations have the following solutions:

\textbf{$\bullet$ $ m>0$.} There is only one solution,  the low energy theory is the TQFT of Acharya and Vafa 
\be
\text{3d} \quad U(k)^{\mathcal{N}=1}_{N-\frac k2, N}.
\ee
(The Chern-Simons level is obtained integrating out the positive mass fermions).

\textbf{$\bullet$ $ m<0$.} There are $k+1$ solutions.  The quarks $XX^{\dag}$ take VEV and break both the flavor symmetry $U(N)\rightarrow U(J)\times U(N-J)$ and the gauge symmetry $U(k)\rightarrow U(k-J)$. The low energy models living on each of the $j$ vacua are 
\be
\label{eq:low energySUNN}
U(k-J)_{-\frac {k-J}2,0}^{\mathcal{N}=1}\times Gr(J,N).
\ee 
The  Chern-Simons levels are  computed looking at the mass of the charged fermions under the unbroken gauge group. All these vacua preserve supersymmetry. At low energies, the $SU(k-J)$ part of the $U(k-J)$ group confines, leaving an $U(1)_0$ (with the exception of $J=k$). This is the signal we were looking for if we were searching for domain walls that break baryonic symmetry. In fact, $U(1)_0\sim S^1$ which is the NLSM expected when baryons take VEV. Moreover, all the low energy models that have an $U(1)_0$ factor have automatically $\text{WI}=0$. The only domain wall contributing to the WI is the one with $J=k$, in which the gauge group is completely broken. This domain wall was already discovered in \cite{Ritz:2004mp}.

Notice that all the NLSM in \eqref{eq:low energySUNN} have a corresponding Wess-Zumino term that can be specified describing the NLSM as a $\cN=1$ $U(k-J)_{\frac{N-J}{2},N-\frac J2}$ and $N$ fundamental scalar multiplets taking VEV.

In summary, the various vacua are listed in Table \ref{tab: solution 3d solution SUNN}.
\begin{table}
 \begin{center}
\begin{tabular}{|c|l|c|}
\hline
Wall&Effective theory& Witten Index\\
\hline
$k$&$U(1)_{0}^{\mathcal{N}=1}\times Gr(J,N)$,\,\, $J=0,\dots,k-1$ & $0$\\
&$Gr(k,N)$&$\smqty(N\\k)$\\
\hline
\end{tabular}
\end{center}
\caption{\label{tab: solution 3d solution SUNN} IR models on the vacuum structure of \eqref{eq:3dtheory} for $m<0$ and $k<\frac N2$. The various contributions to the Witten index are displayed. }
\end{table}
As one can see the two tables, Table \ref{tab: solution 3d solution SUNN} and Table \ref{tab:solution SU N}, match.

We now want to stress that the analysis of the vacua we just perform seems legitimate when $k>\frac N2$. Therefore it seems that for domain walls with $k>\frac N2$, the domain walls solutions should be $k+1$. However from $4d$ parity invariance and the $3d$ $U(k) \leftrightarrow U(N-k)$ duality, we know that for $k>\frac N2$ there must be $N-k+1$ solutions. The full amount vacua of the models with $k>\frac N2$, are $U(1)_0\times \mathrm{Gr}(J,N)$, with $J=0,\dots,k.$ The first $N-k+1$ vacua map to the vacua found above in the dual model. However, the semiclassical analysis yields additional vacua, those from $J=N-k+1$ to $J=k$. 

We ascribe the mismatch to the fact that this interpretation of the analysis is naive and does not consider the fact that strong coupling effects that may arise due to the smallness of the CS level compared to the rank of the gauge algebra. A possible scenario is that our model could have two phase transitions, say for $m=0$ and for $m=m^{\ast}$. The phase of the model when $m<0$ yields $k+1$ vacua, the wrong vacuum structure according to the $4d$ analysis. Instead, when $0<m<m^{\ast}$ the model has $N-k+1$ vacua, the correct vacua matching our $4d$ computation. This means that the transition on the walls is captured by the phase transition around the $m=m^{\ast}$ and not around $m=0$. The phase transition around $m=m^{\ast}$ is not seem by the semiclassical analysis. We leave an investigation of this proposal to future work.

\subsubsection*{The special case of $SU(2) \simeq Sp(1)$}
If $N=2$, the $3d$ model, in the regime corresponding to the 4d description with the constrained Wess- Zumino model $m_{3d}<0$, has two vacua:  $U(1)_0$ and a NLSM with target space $\mathbb{CP}^1$. From the $3d$ $U(1)$-gauge theory  perspective these seem two disconnected sets of vacua, however, because from the $4d$ perspective the $SU(2)$ with $F=2$ model is exactly the same as $Sp(1)$ with $F=2$, we know there should be only one solution, with a NLSM with target space $Sp(2)/Sp(1)\times Sp(1)$, found in \cite{Benvenuti:2021yqv}. 

The interpretation of this apparent mismatch is as follows: the $3d$ $U(1)$-gauge theory has UV global symmetry $U(2)$, but IR global symmetry $Sp(2)$.  Hence when we compute naively (using the UV global symmetry) the NLSM's living on the vacua in the $3d$ $U(1)$-gauge theory, we get a wrong result. The two sets of disconnected vacua we naively see from the UV $U(1)$ perspective are in fact submanifolds of a bigger connected set of vacua, which is instead seen semiclassically in the $3d$ $Sp(1)$-gauge theory discussed in \cite{Benvenuti:2021yqv}, for which the UV and IR symmetry are the same, namely $Sp(2)$.

%%%%%%%%%%%%%%%%%%%%%%%%%%%%%%%%%%%%%%%%%%%%%%%%%%%%%%%%%%%%%%%%%%%%%%%%%%%%%%%%%%%%%%%%%%%%%%%%%%%%%%%%%%%%%%%%%%%%%%%%%%%%%%%%%%%%%%%%%%%%%%%%%%%%%%%%%%%%%%%%%%%%%%%%%%%%%%%%%%%%%%%%%%%%%%%%%%%%%%%%%%%%%%%%%%%%%%%%%%%%%%%%%%%%%%%%%%%%%%%%

\section{BPS domain walls of $SU(N)$ with $N+1$ flavors}\label{secF=N+1}

Now we consider $\cN=1$ $SU(N)$ gauge theory with $F=N+1$ flavors, that means $N+1$ chiral fields in the fundamental $Q^I_a$ and anti-fundamental  $\tilde{Q}_J^a$ representations,   $I,J=1,\dots,N+1$ and $a=1,\dots,N$.
 As is well known, the massless $\cN=1$ $SU(N)$ theory with $N+1$ flavors is described at low energy by a Wess-Zumino model with fields $M_J^I \lra \Tr(Q^I \tilde{Q}_J)$, $B_I \lra \epsilon_{I I_1 \dots I_N}Q_{I_1}\ldots Q_{I_N}$, $\tilde{B}^J \lra \epsilon^{J J_1\ldots J_N}\tilde{Q}_{J_1}\ldots \tilde{Q}_{J_N}$.  The superpotential of the massless Wess-Zumino model is
\be
\label{eq:wessZumino}
W=\frac{1}{\Lambda^{2N-1}}\bigl(B_I M_J^I \tilde{B}^J -  \det(M)\bigr).
\ee   
This superpotential, which is a purely quantum expression (note that classically the rank of $M$ should be $\text{rk}(M)\le N$, giving us $\det(M)=0$), generates the correct moduli space of the massless theory. This moduli space is parametrized by the meson and the baryons of the gauge model, which are related to $M_I^J$, $B_I$, $\tilde{B}^J$. Classically, there are also constraints between baryons and mesons which are precisely the F-term equations derived from the superpotential \eqref{eq:wessZumino}.

Once we introduce a mass term $\delta W = m_{4d} \Tr M$, the moduli space of the massless theory is lifted and the $N$ supersymmetric  vacua become 
\be\label{vacua}
B_I=\tilde{B}^J=0, \quad M_I^J=\omega \, m_{4d}^{\frac{1}{N}} \, \Lambda^{\frac{2N-1}{N}} \, \mathbb{I}_{N+1}, \qquad \text{where } \,\, \omega^N=1 .
\ee
Here $m_{4d}$ is small because \eqref{eq:wessZumino} describes the low energies behavior of SQCD. The WZ is weakly coupled, therefore we use the canonical Kähler potential in terms of  the fields $B_I$, $\tilde{B}^J$ and $M_{J}^I$: $\cK= M_I^J (M^{\dagger})_J^I + B^I B^{\dag}_I +\tilde{B}_J \tilde{B}^{\dag J}$. 

\subsection{Analysis of the BPS equations}
We now study the BPS equations outlined at the beginning of Section \ref{BPS1}. The analysis is similar to the one for $Sp(N)$ with $F=N+2$ flavors studied in \cite{Benvenuti:2021yqv}, the main difference being the presence of the baryons. Let us start with the following observation. Away from the origin of the mesons space, the meson matrix takes a VEV, making all the fields $B_I$, $\tilde{B}^J$ and $M_{J}^I$ massive. From \eqref{eq:wessZumino}, the mass of the $B_I$ and $\tilde{B}^J$ is proportional to $m_{4d}^{\frac 1N}$, while the mass of $M_{I}^J$ is proportional to $m_{4d}$. Since $m_{4d}$ is small, the mass of $B_I$, $\tilde{B}^J$ is much larger than the mass of the $M_{I}^J$. This means that there should not be any solutions of the differential equations where the baryons have a non-zero profile.\footnote{One possible exception to this argument is when the mesonic trajectory of the wall passes through the origin. This will be important in Sec. \ref{parityinv}, where we consider parity invariant walls, for $k=\frac{N}{2}$, that do pass through the origin.} 

Since the fields $B_I$ and $\tilde{B}^J$ are much heavier,  we can integrate them out, obtaining the reduced superpotential
\be
\label{SuperSUN+1}
W=-\frac {1}{\Lambda^{2N-1}}\det(M) + m_{4d} \Tr M.
\ee 
After the diagonalization of the matrix $M$ using the flavor symmetry, the superpotential \eqref{SuperSUN+1} gives us the same differential equations of the superpotential of $Sp(N)$ with $F=N+2$ flavors studied in \cite{Benvenuti:2021yqv}. So the solutions are the same. The difference is that the flavor group symmetry is $U(F)$ instead of $Sp(F)$, which translates in the change of the moduli of the domain wall from the quaternionic Grassmaniann (denoted $\mathrm{HGr}(J,F)$) to the complex Grassmaniann (denoted $\mathrm{Gr}(J,F)$). The list of solutions we have found is in Table \ref{table DW theories SU(N) F=N+1} and are in one-to-one correspondence with the solutions found for $Sp(N)$ with $F=N+2$ of \cite{Benvenuti:2021yqv}.

 \begin{table}
\begin{center}
\begin{tabular}{|c|c|c|}
\hline
Wall&Effective theory& Witten Index\\
\hline
$k$& $\mathrm{Gr}(J,N+1),\quad J\in \{0,\dots,k\}$&$  \mqty(N\\k)=\sum_{j=0}^{k}(-1)^{j+k}\mqty(N+1\\j)$\\
\hline
\end{tabular}
\end{center}
\caption{\label{table DW theories SU(N) F=N+1}Domain wall solutions found for 4d $\cN=1$ $SU(N)$ SQCD with $F=N+1$ flavors in the regime when $m_{4d}\ll\Lambda$. For each $k<\frac N2$-wall sector are included also the various contributions to the Witten index from each solution.}
\end{table}

Each $k$-wall sector consists of $k+1$ different solutions, parameterized by the integer $J=0,1,\ldots,k$. Each wall hosts a trivial a TQFT and a non-trivial NLSM, with target space
\be \mathrm{Gr}(J,F) = \frac{U(F)}{U(J) \times U(F-J)} \,,\ee
the flavor symmetry being broken as $U(F) \ra U(J) \times U(F-J)$.

\begin{figure}[h!]
\centering
\hspace{\stretch{1}} \includegraphics[width=.32\textwidth]{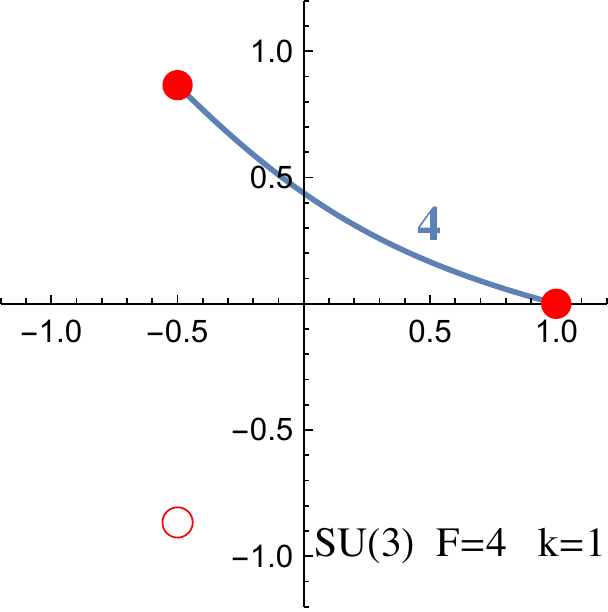}\hspace{\stretch{1}}
 \includegraphics[width=.32\textwidth]{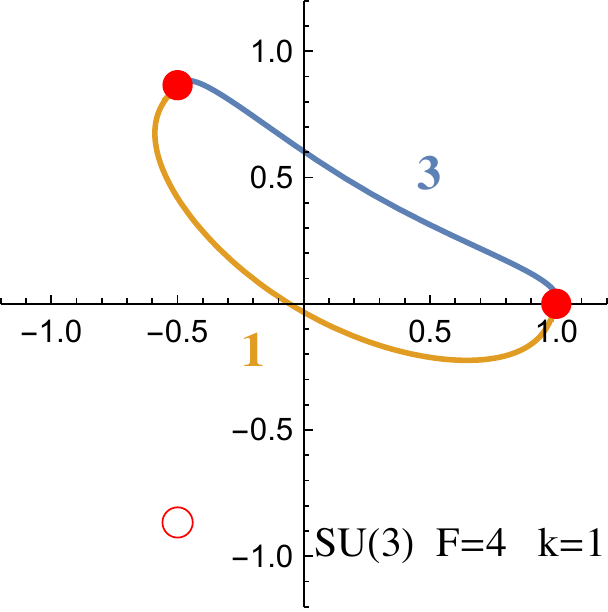}\hspace{\stretch{1}}
 \caption{Example of  1-wall solutions for $SU(3)$ with $F=4$ flavors.}
 \label{fig:wallsu3}
\end{figure}

\begin{figure}[h!]
\centering
\hspace{\stretch{1}} \includegraphics[width=.32\textwidth]{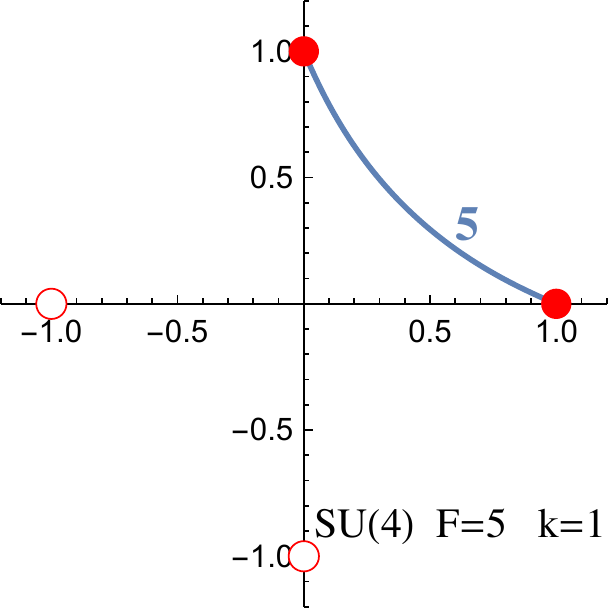}\hspace{\stretch{1}}
 \includegraphics[width=.32\textwidth]{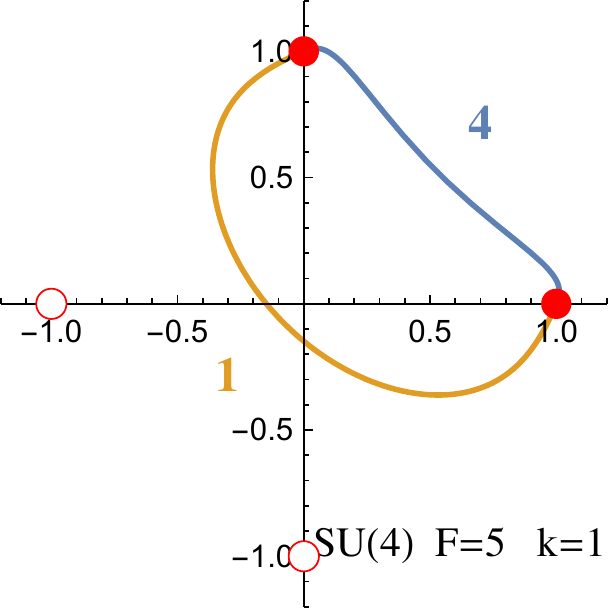}\hspace{\stretch{1}}
 \includegraphics[width=.32\textwidth]{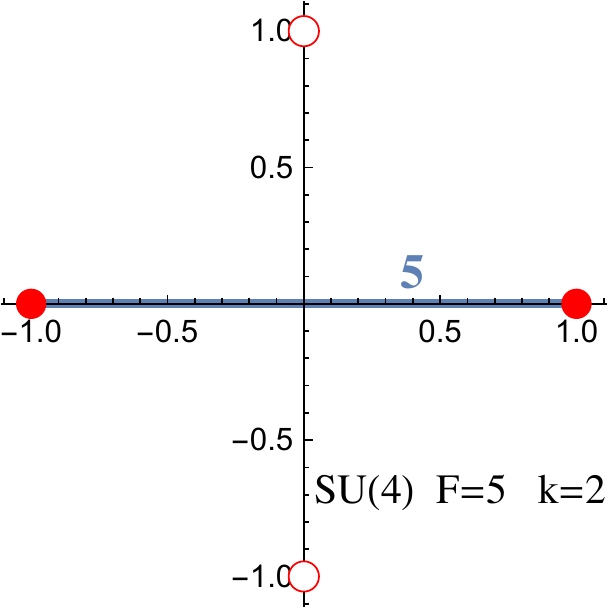}\hspace{\stretch{1}}
 \caption{Example of 1-wall solutions and 2-wall solutions for $SU(4)$ with $F=5$ flavors.}
 \label{fig:wallsu4}
\end{figure}

\begin{figure}[h!]
\centering
\hspace{\stretch{1}} \includegraphics[width=.32\textwidth]{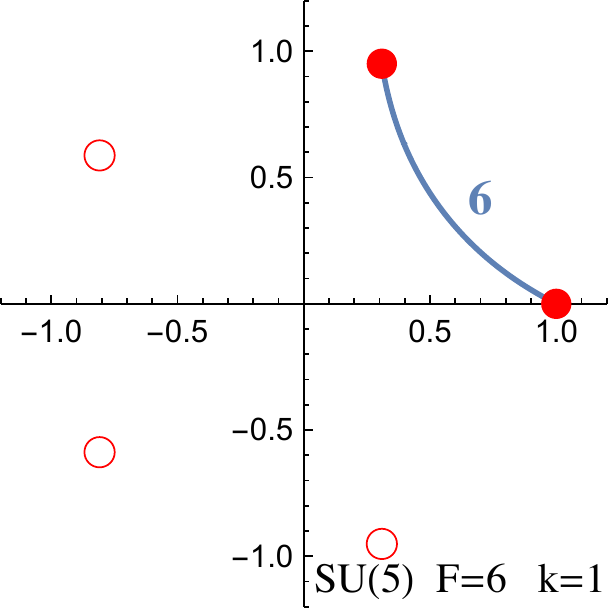}\hspace{\stretch{1}}
 \includegraphics[width=.32\textwidth]{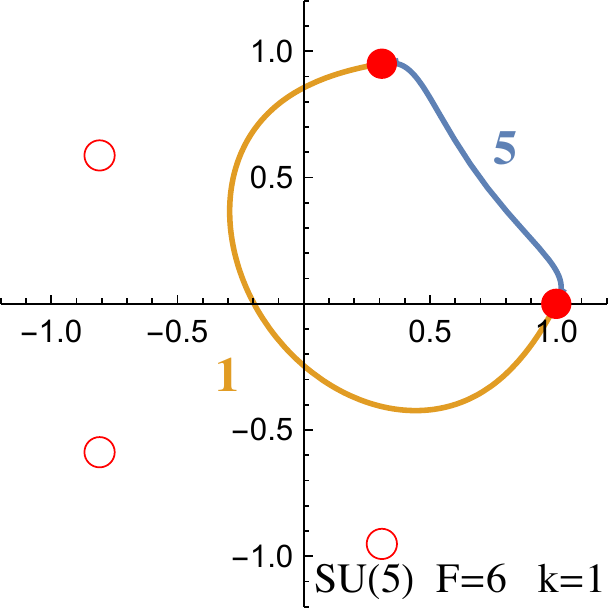}\hspace{\stretch{1}}\\
\hspace{\stretch{1}} \includegraphics[width=.32\textwidth]{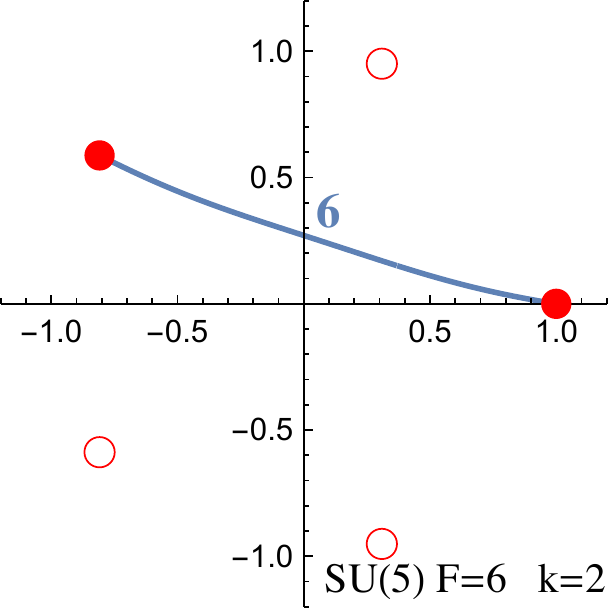}\hspace{\stretch{1}}
 \includegraphics[width=.32\textwidth]{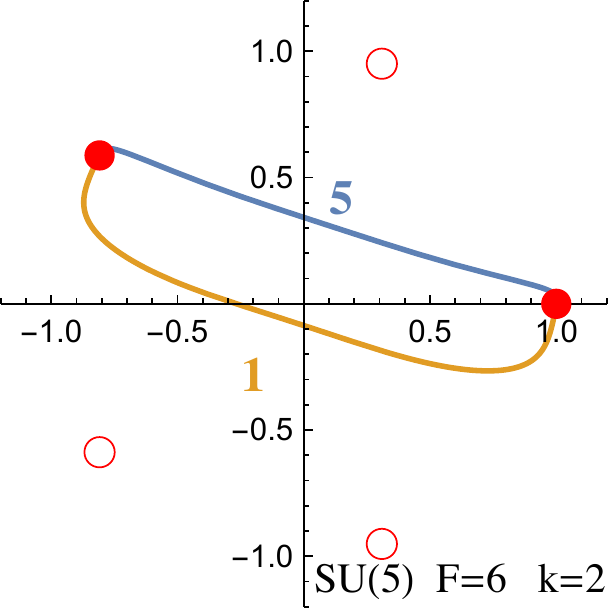}\hspace{\stretch{1}}
 \includegraphics[width=.32\textwidth]{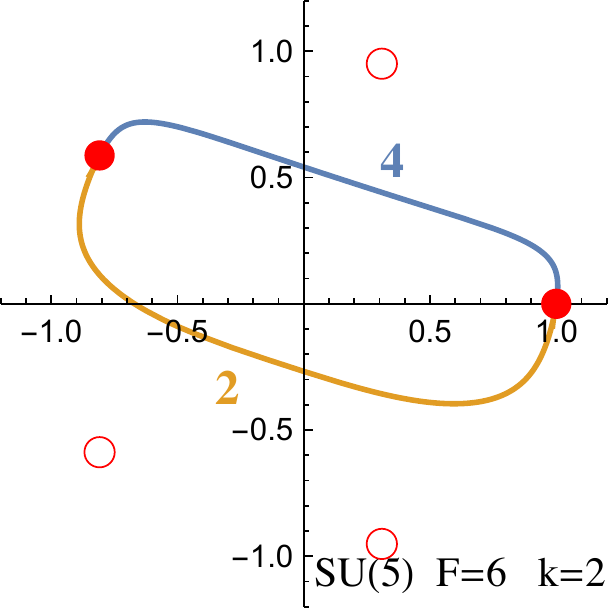}\hspace{\stretch{1}}
 \caption{Example of 1-wall  and 2-wall solutions for $SU(5)$ with $F=6$ flavors.}
 \label{fig:wallsu5}
\end{figure}

We display the trajectories of the mesonic eigenvalues in Figures \ref{fig:wallsu3}, \ref{fig:wallsu4} and \ref{fig:wallsu5}.

\label{sec: SQCDN+1}

\subsubsection{The parity-invariant walls}\label{parityinv}
Similar to what happens for $Sp(N)$ with $N+2$ flavors \cite{Benvenuti:2021yqv}, the parity-invariant walls of $SU(N)$ with $N+1$ flavors, that is the $k=\frac N2$ when $N$ is even, require a special treatment. 

In this case, the numerical analysis of \eqref{eq:diff} yields a single domain wall that connects the two vacua $M=\mathbb{I}_{N+1}$ and  $M=-\mathbb{I}_{N+1}$ (here we have rescaled $M$ in units of $m_{4d}^{\frac{1}{N}}\Lambda^{\frac{2N-1}{N}}$) along the real line, hence passing through the origin, see right picture in Figure \ref{fig:wallsu4}. Having a single solution, invariant under the global symmetry, is in contrast with the expectations coming from $k \neq \frac N2$, where we find $k+1$ domain walls. 

In \cite{Benvenuti:2021yqv} we showed that the single naive solution of the parity invariant wall must be interpreted as the coalescence of many different solutions. The strategy was to deform the Kähler potential. Upon making an infinitesimal deformation of the Kähler potential, more solutions appear. Such deformations, however, break the flavor symmetry, so there is no automatic recipe to obtain the full moduli space of solutions. The saturation of the Witten index is also problematic.

The solutions found in \cite{Benvenuti:2021yqv} for $Sp(N-1)$ with $N+1$ flavors carry over to $SU(N)$ with $N+1$ flavors, so we do not repeat the same analysis here.

One last comment about possible baryonic solutions. Since the solutions pass through the origin of the mesonic space, the argument given before \eqref{SuperSUN+1} does not apply here. As we show below in the special case of $SU(2) = Sp(1)$, upon deforming the Kähler potential, there are baryonic solutions.

\begin{figure}[h!]
\centering
\hspace{\stretch{1}} \includegraphics[width=.32\textwidth]{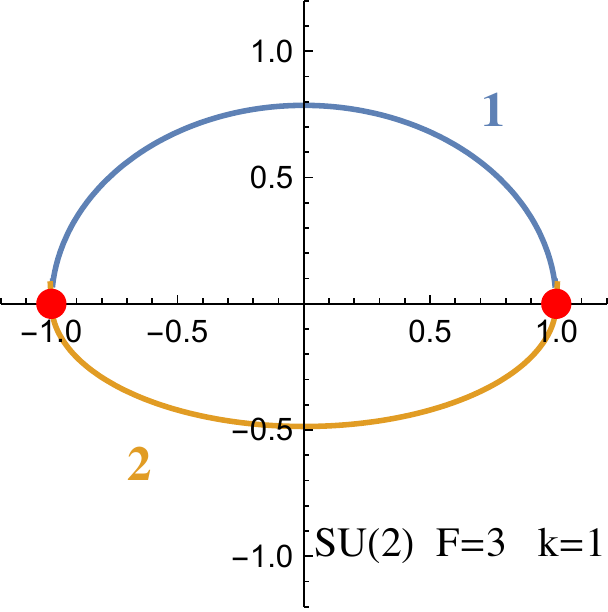}\hspace{\stretch{1}}
 \includegraphics[width=.32\textwidth]{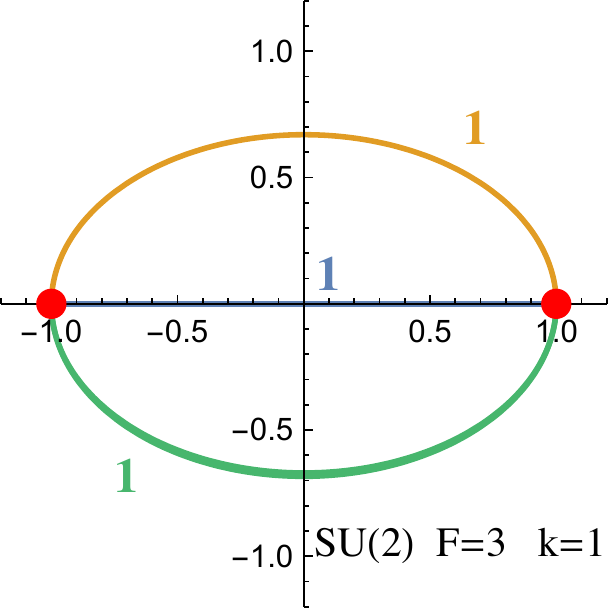}\hspace{\stretch{1}}
 \caption{Example of 1-wall   solutions for $SU(2)$ with $F=3$ flavors.}
 \label{fig:1wallsu2}
\end{figure}

For $SU(2)$ with $3$ flavors, which has global symmetry $Sp(3)$, one of the deformations studied in  \cite{Benvenuti:2021yqv} is
\be
 \delta \cK=\frac14\Tr(MJ)\Tr(M^{\ast}J)\,,
\ee
where $J=i\sigma_2\otimes \text{diag}(0.1,0.1,0.0)$. This deformation breaks explicitly the flavor symmetry $Sp(3)\rightarrow Sp(2)\times Sp(1)$. The solutions we have found after such deformation are in Figure \ref{fig:1wallsu2}. 
\begin{figure}[h!]
\centering
\hspace{\stretch{1}} \includegraphics[width=.32\textwidth]{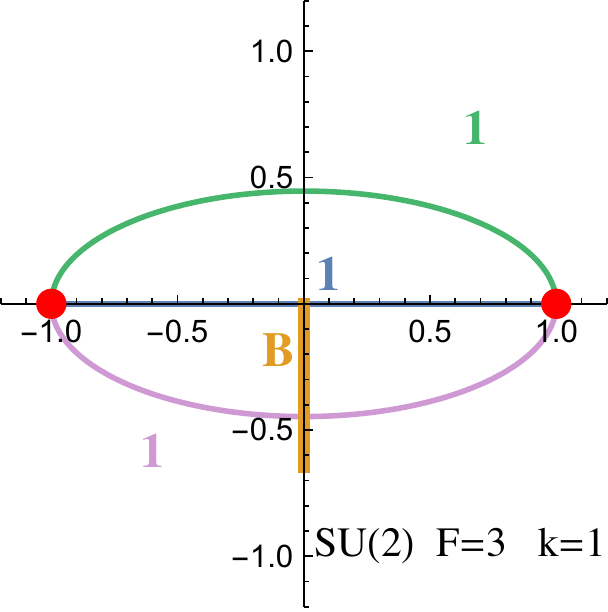}\hspace{\stretch{1}}
 \caption{Example of 1-wall  solution for $SU(2)$ with $F=3$ flavors, where the baryons have a non zero profile. We used different scales to plot these trajectories: the imaginary axis is zoomed in $10 \times $.}
 \label{fig:1wallsu2B}
\end{figure}
 We  see that after the deformation a solution involving the baryon is possible.  Indeed, using the residual flavor symmetry $Sp(2)$ we can rotate the figure on the right into the Figure \ref{fig:1wallsu2B} where a baryon is different from zero.  One can see this using the map between the operators in the $SU(2)$ language and the $Sp(1)$ language 
\be
M_{Sp(1)}=\begin{pmatrix}
0&q_3&-q_2&M_{11}&M_{12}&M_{13}\\
-q_3&0&q_1&M_{21}&M_{22}&M_{23}\\
q_2&-q_1&0&M_{31}&M_{32}&M_{33}\\
-M_{11}&-M_{21}&-M_{31}&0&\tilde{q}_3&-\tilde{q}_2\\
-M_{12}&-M_{22}&-M_{32}&-\tilde{q}_3&0&\tilde{q}_1\\
-M_{13}&-M_{23}&-M_{33}&\tilde{q}_2&-\tilde{q}_1&0
\end{pmatrix}
\ee
and the rotation matrix
\be
T=\mqty(
1&0&0&0&1&0\\
0&1&0&1&0&0\\
0&0&1&0&0&0\\
0&\frac12&0&\frac32&0&0\\
\frac12&0&0&0&\frac32&0\\
0&0&0&0&0&1
)\in Sp(2)\times Sp(1)
\ee

%%%%%%%%%%%%%%%%%%%%%%%%%%%%%%%%%%%%%%%%%%%%%%%%%%%%%%%%%%%%%%%%%%%%%%%%%%%%%%%%%%%%%%%%%%%%%%%%%%%%%%%%%%%%%%%%%%%%%%%%%%%%%%%%%%%%%%%%%%%%%%%%%%%%%%%%%%%%%%%%%%%%%%%%%%%%%%%%%%%%%%%%%%%%%%%%%%%%%%%%%%%%%%%%%%%%%%%%%%%%%%%%%%%%%%%%%%%%%
\subsection{Effective $3d$ theory on the walls: monopole superpotential}\label{SUN+1}

We  now turn to our proposal for the $3d$ effective theory on the $k$-domain wall of  $4d$ $SU(N)$ with $N+1$ flavors. Our proposal is inspired by the following $3d$ $\cN=2$ duality \cite{Nii:2020xgd,Amariti:2020xqm}:

\be \label{Nii} \ba{c} SU(N)^{\cN=2}_{-(N+1)/2} \, \text{w/ $(0,N+1)$ flavors $Q_i$} \\ \cW_{\cN=2}=0 \ea
\Llra
\ba{c} U(1)^{\cN=2}_{N/2} \, \text{w/ $((N+1)_{q_i}, 1_{\tilde{p}})$ flavors} \\ \cW_{\cN=2}= \M_{BPS}^+ \ea \ee
The $SU(N)$ theory on the l.h.s. is expected to be related to the interface theory of $4d$ $SU(N)$ with $N+1$ flavors, so we can also expect that the $U(1)$ theory on the r.h.s. is the $\cN=2$ ancestor of the $3d$ theory living on the 1-wall. In other words we are extending the $1$-wall $\lra$ interface duality \ref{SUU1f}  to the case of $F=N+1$. Notice that in the $U(1)$ theory there is a superpotential term linear in the supersymmetric monopole, denoted $\M_{BPS}^+$. This is a qualitatively new feature, compared to previously discussed cases.

The CS matter theory admits a supersymmetric monopole because the difference between the number of anti-fundamentals minus the number of fundamentals is precisely twice the CS level. The chiral ring generators in the $SU(N)$ theory are the $N+1$ baryons $\varepsilon_N Q^N$. These baryons are mapped to the $N+1$ mesons $\tilde{p} q_i$ in the dual $U(1)$ theory.

Theories with unitary gauge groups and monopole superpotentials were studied in \cite{Benini:2017dud}. Duality $8.18$-$8.19$ of \cite{Benini:2017dud} (setting $N_c=k$, $N_f=N+1$, $\kappa=N$) reads

\be\label{1Mdual} \ba{c}  U(k)^{\cN=2}_{N/2}  \, \text{w/ $(N+1, 1)$ flavors} \\ \cW_{\cN=2}=\M_{BPS}^+ \ea
       \Llra
\ba{c} U(N-k)^{\cN=2}_{-N/2}   \, \text{w/ $(1_P,(N+1)_{\tilde{Q}_i})$ flavors} \\ \cW_{\cN=2}=\mu_i \, tr(P \tilde{Q}_i ) +\M_{BPS}^- \ea \ee

The chiral ring is generated by the $N+1$ mesons on the l.h.s., which are mapped to the gauge singlets $\mu_i$. The continuos flavor symmetry is $SU(N+1) \times U(1)$ (without the monopole superpotential there would be an additional $U(1)$ factor, the monopole superpotential breaks a combination of the axial and the topological, or magnetic, symmetry).

Because of the two dualities just presented, we expect the $\cN=2$ theories in \ref{1Mdual} to be related to the $k$-wall of $4d$ $SU(N)$ with $N+1$ flavors. The domain wall is described by an  $\cN=1$ SCFT obtained with an $\cN=1$ perturbation of the above $\cN=2$ fixed points, that, among other things, gives mass to the gauge singlet fields $\mu_i$. It is important that the monopole term in the superpotential survives the deformation. More precisely in $\cN=1$ language, such term is a dressed monopole of the form
\be  \M^+_{dressed} = \M^+_{bare} \, p \, q_1 q_2 \ldots q_{N+1} \varepsilon_{N+1} \ee
which is gauge-invariant (the flavors $q_i$/$p$ have $+1$/$-1$ charge under the gauge $U(1) \subset U(k)$) and also invariant under the $SU(N+1) \times U(1)$ global symmetry. $\M^+_{bare}$ is defined as a disorder operator, creating a flux $(1,0,\dots,0)$ in the Cartan's of the gauge group of $U(k)$.

Summing up, our proposal for the $3d$ theory living on the $k$-wall is
\be\label{wallSUN+1} \ba{c}  U(k)^{\cN=1}_{(N-k)/2, N/2}  \,\,\, \text{with $N+1$ fundamentals $q_i$ and $1$ fundamental $p$,} \\ \cW =\frac14\Tr(q^\dagger_i q^jq^\dagger_j q^i) + \frac A4 \Tr(q^\dagger_i q^i) ^2 +\frac B4 \Tr(p^\dagger p)^2 
+\frac{\eta}{2}\Tr(q^\dagger_i q^i) \Tr(p^\dagger p) +\\+\frac{\alpha}{2}\Tr(p^\dagger q^\dagger_j) \Tr(q^j p)+ \M_{dressed}^+ \ea \ee

where $A,B,\alpha, \eta$ are real parameters.\footnote{A minor consistency check of our proposal is that we can mass deform the theory \ref{wallSUN+1}  with the mass term 
\be \delta \cW = tr(p \, q_{N+1}) \,,\ee
 breaking the global symmetry as
\be SU(N+1) \times U(1) \rightarrow SU(N) \times U(1) \,.\ee
 The CS level is unchanged (because the real masses of $p$ and $q_{N+1}$ have opposite signs). The monopole superpotential is lifted (otherwise the global symmetry in the IR would only be $SU(N)$). Hence, in the infrared, we end up on the $U(k)_{\frac{N-k}{2},\frac N2}^{\cN=1}$ theory with $N$ flavors which is our proposal for the domain walls of $SU(N)$ with $N$ flavors \eqref{eq:3dtheory}.
} It will turn out that the vacua of the mass-deformed \eqref{wallSUN+1} are related to our $4d$ analysis when $A>0$, $B<0$ and $\eta<-1$. We conjecture that an $\cN=1$ SCFT in this region of parameters exists. It would be nice to study the existence of such fixed point further, testing the validity of our assumptions.

Theory \ref{wallSUN+1} has two independent $SU(F)$-invariant mass terms, namely $\Tr(p^\dagger p)$ and $\Tr(q^\dagger_i q^i)$. In order to interpolate between massive phases which are related to the domain walls we turn on the following combination:
\be \delta \cW = m \left(  \Tr(q^\dagger_i q^i) -\Tr(p^\dagger p) \right) \ee
In this way, for positive $m$, $N+1$ flavors have positive  mass and $1$ flavor has negative  mass, hence the vacuum is described by $U(k)^{\cN=1}_{N-k/2, N}$, reproducing the AV TQFT as required.

\subsubsection*{Analysis of the massive vacua}
In order to analyze the vacua of \ref{wallSUN+1}, we can set $B=-1$ for simplicity and $\alpha=0$ for the purpose of this section\footnote{The general analysis of the vacuum structure of the model is done in the appendix \ref{vacuum:analysis}. For different choices of the parameters there are vacuum structures which seem not to be related to to domain walls of $4d$ $SU(N)$ with $N+1$ flavors.}.

 We diagonalize the matrix $q q^{\dag}=\text{diag}(\lambda_1^2,\dots,\lambda_{N+1}^2)$ using the flavor and gauge symmetry (with $\lambda_i^2\ge0$ and with at most $k$ of them $\lambda_i>0$).

Supersymmetric vacua satisfy the F-term equations
\bea
\lambda_i\bigl(\lambda_i^2+\eta |p|^2+\alpha p_i^2+m\bigr)=0\\
p_a \bigl(B \abs{p}^2+ \eta \sum_j\lambda_j^2+\alpha\lambda_a^2-m\bigr)=0.
\eea

Notice that if both and $q_i$ and $p$ take a vev, the $U(1)$ factor in the global symmetry is broken, hence such vacua should map to would-be-domain walls where the $4d$ baryonic symmetry is broken, and we did not find any such solution in the previous section.

\textbf{$\bullet$ $ m>0$.} There is only one solution, $p = q_i = 0$, the low energy theory is the TQFT 
\be
\text{3d} \quad U(k)^{\mathcal{N}=1}_{N-\frac k2, N}.
\ee
The monopole superpotential of the UV theory has no effect on these IR vacua.

\textbf{$\bullet$ $ m<0$.} The solutions with $p\neq 0$ typically can appear and disappear changing the real parameters of the quartic superpotential, we analyze them in Appendix \ref{vacuum:analysis}. Here we discuss only the solutions with $p=0$, which should be the ones related to domain walls. There are $k+1$ different vacua, parametrized by $J=0,\dots,k$:
\be  p=0, \qquad \lambda_i^2=-\frac{m}{1+J A}\,\,\,\, \text{ for } i=0,\dots, J, \qquad \lambda_i=0\,\, \text{for } i=J+1,\dots, k.\ee
The low energy theory includes a TQFT factor $U(k-J)_{-\frac {k-J}2,0}^{\mathcal{N}=1}$, which without the monopole superpotential would be equivalent to $U(1)_0 \sim S^1$. The monopole superpotential lifts the $S^1$ to two points. This phenomenon occurs because, once we have integrated out all the flavors and the CS level of the $U(1)$ happens to be zero, we can define the gauge invariant monopole as the $\M^+$ as the exponential of the dual photon $\theta$,  $e^{i\theta}$. Therefore the superpotential $\cW=\Re(\M^+)$ can be written as $\cW=\cos(\theta)$. The F-term equations of such superpotential give us two vacua. One can also see this phenomenon studying a QFT for which a convenient duality is known. Such example is explained in Appendix \ref{monopole:superpotential}. Hence the low energy theory is
\be
\label{eq:low energySUNN+1}
\bZ_2 \times Gr(J,N+1).
\ee 
Modulo the double degeneracy, these are the $k+1$ vacua expected from the $4d$ analysis. 

Putting this result together with the $\cN=2$ dualities discussed above, we gathered quite a bit of evidence that the theories \ref{wallSUN+1}, inside an appropriate region of the parameter space, describe the domain walls of $4d$ $SU(N)$ with $N+1$ flavors. It would be nice to test this proposal further.

\section*{Acknowledgements}

We are very grateful to Francesco Benini for initial collaboration, useful suggestions
and careful reading of the manuscript. We also thank Matteo Bertolini for his invaluable suggestions and comments throughout the development of the paper. 

\appendix

\section{Failure of a different $3d$ model for $SU(N)$ with $F=N+1$}\label{sec:DG}
In this Appendix we analyze the straightforward generalization of the models that describe the domain walls of $SU(N)$ with $F \leq N$ flavors, that for $F=N+1$ reads
\be
U(k)_{\frac{N-k-1}{2},\frac {N-1}2}^{\cN=1} \quad \text{with $N+1$ fundamentals } X, \quad \cW \sim + |X|^4.
\label{eq:SUN+1v2}
\ee
This model is the natural extension of the proposal \eqref{eq:3dtheory}, adding one fundamental and changing the CS levels accordingly. The global symmetries meet the requirements. 
%Second, if we restrict ourselves to $k<\frac{N}{2}$ we can see that the vacua of this model almost reproduce the domain wall solutions we have found in $4d$ analysis. 
The vacua analysis follows the same path of Section \ref{SUNeffective}. The full superpotential, including the mass term, is
\be
\cW=\frac{1}{4}\Tr(XX^{\dag}XX^{\dag})+\frac{\alpha}{4}\Tr(XX^{\dag})^2+m \Tr XX^{\dag}
\ee 

The analysis of the vacua is carried out  diagonalizing the matrix $XX^{\dag}=\text{diag}(\lambda_1^2,\dots,\lambda_N^2)$ using the flavor and gauge symmetry. Since $XX^{\dag}$ is semi-positive definite, $\lambda_i^2\ge0$. As usual we set $\alpha=0$. Supersymmetric vacua satisfy the F-term equations
\be
\lambda_i(\lambda_i^2+m)=0\quad  i\in \{0,\dots,k\}.
\ee
These equations have the following solutions:

\textbf{$\bullet$ $ m>0$.} There is only one solution,  the low energy theory is the TQFT of Acharya and Vafa 
\be
\text{3d} \quad U(k)^{\mathcal{N}=1}_{N-\frac k2, N}.
\ee
(The Chern-Simons level is obtained integrating out the positive mass fermions).

\textbf{$\bullet$ $ m<0$.} There are $k+1$ solutions.  The quarks $XX^{\dag}$ take VEV and break both the flavor symmetry $U(N)\rightarrow U(J)\times U(N-J)$ and the gauge symmetry $U(k)\rightarrow U(k-J)$. The low energy models living on each of the $J$ vacua are 
\be
\label{eq:low energySUNN1}
U(k-J)_{-\frac {k-J+2}2,-1}^{\mathcal{N}=1}\times Gr(J,N).
\ee 
Along with the Grassmannian, there is a TQFT: $U(k-J)^{\cN=1}_{-\frac {k-J+2}2,-1}=U(k-J)^{\cN=0}_{-1,-1}$. This topological quantum field theory is almost trivial since it has only one transparent line, which has spin $\frac12$ (see \cite{Hsin:2016blu}). In particular this model has $WI=1$. The presence of the non-trivial TQFT is a problem for the proposal \eqref{eq:SUN+1v2}, since in $4d$ we have a Wess-Zumino model.

%(Here we should point out that for $k\ge \frac N2$ the vacuum analysis fails to reproduce the correct $4d$ solutions found.  This is not the first time we encounter a similar behavior: in \cite{Benvenuti:2021yqv} we saw exactly the same phenomenon in the $Sp(N)$ with $F=N+2$ domain wall case.) 
%There is however a substantial difference with what happened in \cite{Benvenuti:2021yqv}. In that case we have argued that there still is a duality between the theories on the $k$-walls and $N+1-k$-walls\footnote{Pay attention that the analogous of the duality relation $k \leftrightarrow N-k$ in the $SU(N)$ case is $k\leftrightarrow N+1-k$ for $Sp(N)$}, arguing that the mismatch in the vacuum analysis was due to a naive analysis when the CS level was too small compared to the rank of the $3d$ gauge group. 

The proposal \eqref{eq:SUN+1v2} also has another problem, related to dualities. For all the domain walls studied ($Sp(N)$ with $F\leq N+2$ and $SU(N)$ with $F \leq N+1$), there is always a non trivial  duality incarnating the equivalence between a $k$-wall and the $h-k$ parity reversed wall. Here, however there is not  a $\cN=2$ duality from which one can hope to derive such an $\cN=1$ duality. Indeed, the $\cN=2$ duality satisfied by the $\cN=2$ cousin of \eqref{eq:SUN+1v2} is \cite{Benini:2011mf}
\be\label{n=2SUdualityF=N+1}
 \ba{c} U(k)^{\cN=2}_{\frac{N-1}{2}} \,\,\,\,\, \text{w/ $(N+1,0)$ flavors} \\ \cW_{\cN=2}=0 \ea
       \Llra
\ba{c} U(N+1-k)^{\cN=2}_{-\frac{N-1}{2}} \,\,\,\,\, \text{w/ $(0,N+1)$ flavors} \\ \cW_{\cN=2}=\mu \, \M \ea \ee
%This $\cN=2$ duality does not send the $k$-wall model into the $N-k$-wall model as expected from the $4d$ perspective. 

Deforming this $\cN=2$ duality, we get the $\cN=1$ duality enjoyed by the proposal \eqref{eq:SUN+1v2}, which is of the form 
\be U(k)_{\frac{N-k-1}{2},\frac {N-1}2}^{\cN=1}\,\, \text{with $N+1$ fundamentals} \quad \lra 
\quad U(N+1-k)_{\frac{2-k}{2},-\frac{N-1}{2}}^{\cN=1}\,\, \text{with $N+1$ fundamentals} \,.\ee
This is not the duality which must be enjoyed by the $k$-wall of $SU(N)$ with $N+1$ flavors. Indeed the correct theory must enjoy a duality of the form $U(k)_{*, *}^{\cN=1} \lra U(N-k)_{*,*}^{\cN=1}$. We conclude that the model \eqref{eq:SUN+1v2} cannot describe the $k$-wall of $SU(N)$ with $N+1$ flavors.

\section{Vacuum structure analysis for the 3d domain wall theory of $SU(N)$ with $N+1$ flavors}\label{vacuum:analysis}

In this appendix we are going to study the vacuum structure of the model \eqref{wallSUN+1} in full generality. To easy the reading we report here the superpotential of \eqref{wallSUN+1}
\begin{multline}
\cW =\frac14\Tr(q^\dagger_i q^jq^\dagger_j q^i) + \frac A4 \Tr(q^\dagger_i q^i) ^2 +\frac B4 \Tr(p^\dagger p)^2 +\frac{\eta}{2}\Tr(q^\dagger_i q^i) \Tr(p^\dagger p) +\\+
\frac{\alpha}{2}\Tr(p^\dagger q^\dagger_j) \Tr(q^j p)+ \M_{dressed}^+.
\end{multline}

Using the flavor and the gauge symmetry we can put in a semi-diagonal form the matrix $q^i_a=\mqty(L_{k\times k} &Z_{k\times F-k})$, where $L=\text{diag}(\lambda_1,\dots,\lambda_k)$ and $Z$ has zeros in all the entries. Since $q^\dagger_i q^i$ is positive definite, $\lambda_i\ge0$. Having performed these simplifications, the F-term equations are:
\bea
\lambda_i\bigl(\lambda_i^2+A \sum_j \lambda_j^2+\eta \abs{p}^2+\alpha \abs{p_i}^2+m\bigr)=0\\
p_a \bigl(B \abs{p}^2+ \eta \sum_j\lambda_j^2+\alpha\lambda_a^2-m\bigr)=0.
\eea

Compiuting the second derivative of the superpotential we get the fermion mass matrix. This matrix is important because the in order to know which is the shift of the CS terms due to integrating out massive fermions, we need to know the sign of the fermion masses. The mass matrix reads
\bea
&\frac{\partial \cW}{\partial \lambda_i^2} =\lambda_i^2+A\sum_j\lambda_j^2+\eta \abs{p}^2+ \alpha \abs{p_i}^2+m +2\lambda_i^2(1+A)\\
&\frac{\partial \cW}{\partial \lambda_i\partial\lambda_j}=2A \lambda_i\lambda_j\\
&\frac{\partial \cW}{\partial \lambda_i\partial p_a}=\lambda_i(\eta p^{\ast}_a+2\alpha \delta_i^a p_a^{\ast})\\
&\frac{\partial \cW}{\partial p_a^{\ast}\partial p_b}=\delta_{ab}(B\abs{p}^2+\eta \sum_j \lambda_j^2+\alpha\lambda_a^2-m)+Bp^{\ast}_bp_a
\eea  

The list of solutions for generic $A,B,\eta,\alpha$ and $m$ are the following.

\begin{enumerate}

\item The trivial solution in which $\lambda_i=p_i=0$ is always present regardless the various coefficients of the superpotential. For positive masses it gives us the AV phase, while for negative masses gives us two vacua due to the presence of the monopole in the superpotential which lifts the remaining $S^1$.

\item Another solution is given by
\be
\lambda_i^2=0\quad \text{and} \quad\abs{p}^2=\frac{m}{B}.
\ee
If we want that the only vacuum for positive masses is the trivial one we need to assume that $B<0$. Under this assumption, we get at low energy a TQFT $U(k-1)_{-k/2+1/2,0}$ which is equivalent to $U(1)_0 \sim S^1$, lifted to two points by the monopole superpotential. The existence of this vacuum it is trouble for us. It is not seen in the 4d analysis, yet it is found semiclassically.  If $\frac{\eta+\alpha}{B}<-1$, which changes the masses of the charged fermions, we have a TQFT $U(k-1)_{N-(k-1)/2,0}$, which is not related to the $4d$ analysis.

\item The next set of $k$ vacua are given by the solution
\be
\lambda_i^2=-\frac{m}{(1+J A)}\quad  \text{and}  \quad p_i=0, \quad J=1,\dots,k,
\ee
Let us call these vacua mesonic $J$-vacua. If $\frac{\eta+\alpha}{1+J A}<1$ the low energy theory includes a TQFT factor $U(k-J)_{-\frac {k-J}2,0}^{\mathcal{N}=1}$, which is equivalent to $U(1)_0 \sim S^1$. The monopole superpotential lifts the $S^1$ to two points. Hence the low energy theory is
\be
\label{eq:low energySUNN+1}
\bZ_2 \times Gr(J,N+1).
\ee 
Modulo the double degeneracy, these are the $k+1$ vacua expected from the $4d$ analysis. If $\frac{\eta+\alpha}{1+J A}>1$, at low energy we get a non-trivial  TQFT $U(k-J)_{-(k-J+2)/2,0}$ which seems not to be related to the $4d$ analysis.

\item The next set of $k$ vacua are given by
\be
\label{baryonic:vacua}
\lambda_i^2=-\frac{m (B + \eta)}{(B + J A B - J \eta^2)}\quad  \text{and} \quad\abs{p}^2=\frac{m (1 + JA + J \eta)}{(B + JAB-J \eta^2)}.
\ee
Let us call these vacua baryonic $J$-vacua. Such vacua are not seen in the $4d$ analysis. So since we have assumed that $B<0$, they exist only if small $\eta$, that is $B+\eta$ and  $1+A J+ J\eta$ have opposite  sign. Therefore in order to discard such solutions we assume that at our fixed point $\abs{\eta}$ is big enough.

\item  Last, we have other $k(k-1)$ vacua, which are parametrized by $J,H=1,\dots,k$ and $J\ge H$, given by
\bea
&\lambda_i^2=-m\frac{J(B+\eta)+\alpha(1+(J-H)(A+\eta))}{D},\quad \\
&\quad \abs{p}^2= m H \frac{1+\alpha +J (A+\eta)}{D},\\
&D=(H(1+JA)B-(\alpha+H \eta)^2-(J-H)(A \alpha^2+H\eta^2)).
\eea
These vacua exists on if $J(B+\eta)+\alpha(1+(J-H)(A+\eta))$ and $1+\alpha +J (A+\eta)$ have opposite sign. Needless to say that for our purposes it is easy to tune the parameter $\alpha$ to assure the discarding of such solutions. For example, if $\alpha=0$ we see that, provided the other parameters have been chosen so that the baryonic vacua \eqref{baryonic:vacua} are not present, are automatically not there.
\end{enumerate}

\section{sQED with a linear monopole superpotential}\label{monopole:superpotential}

In this short appendix we analyze the vacuum structure of the 3d model $\cN=1$ $U(1)_{\frac{1}{2}}$ with one charge-$1$ chiral $Q$. This model is well suited to understand the effects of a deformation with a linear monopole superpotential term $\delta\cW=\Re(\M)$.  This theory has a dual model which has been described in \cite{Benini:2018umh}. The duality we are considering is 
\be\label{sQED:monopole} \ba{c} U(1)^{\cN=1}_{\frac12} \, \text{with one chiral } Q \\ \cW=-\frac 14\abs{Q}^4 \ea
\Llra
\ba{c} \text{WZ model} \, \text{with a real } H \text{ and a complex  } P \text{ superfields} \\ \cW= H\abs{P}^2-\frac{H^3}{3} \ea \ee
The basic operator map is
\be
\left\{
\begin{aligned}
&\M\\
&\abs{Q}^2
\end{aligned}
\right\}
\Llra
\left\{
\begin{aligned}
&P\\
&H
\end{aligned}
\right\},
\ee
where $\M$ is the dressed monopole operator of the sQED. We can check that the duality \eqref{sQED:monopole} is a sound proposal, studying the massive vacua of both models. If we deform sQED with a mass term $\delta\cW=m\abs{Q}^2$, we get that the F-term equation  is $Q(-\abs{Q}^2+m)=0$. This in turn gives us the vacuum structure:
\begin{itemize}
\item $m>0$ there are two vacua $Q=0$ and $\abs{Q}=m$. Both vacua support a gapped theory, one because of the triviality of the TQFT $U(1)_1$ and the other after the Higgs mechanism.
\item $m<0$ these is only one vacuum $Q=0$. The low energy model is a $U(1)_0$ which can be dualized in a NLSM with target space $S^1$. 
\end{itemize}

In the dual model the mass deformation maps into $\delta\cW=mH$ and the F-term equations become $\abs{P}^2-H^2+m=0$ and $PH=0$. The vacuum structure of the model is: 
\begin{itemize}
\item $m>0$ $H=\pm\sqrt{m}$ and $P=0$. These two solutions support a gapped vacua which match the $m>0$ phase of sQED.
\item $m<0$ $H=0$ and $\abs{P}^2=-m$. The global $U(1)$ symmetry is broken by the VEV of the complex superfield $P$ giving us a Goldstone boson which lives in $S^1$, matching the $m<0$ phase of sQED.
\end{itemize}

Here we are interested in deformed the above sQED model with $\delta\cW=\Re(\M)$, which maps to $\delta\cW = \Re( P)$ in the dual Wess-Zumino model.  So we want to study the phases varying the parameter $m$ of the Wess--Zumino model with superpotential
 \be
 \cW= H\abs{P}^2-\frac{H^3}{3}+\frac{P+P^{\dagger}}{2}+mH.
 \ee   
 The F-term equations are
 \bea
 &\abs{P}^2-H^2+m=0,\\
 &HP+\frac12=0,\quad\text{and}\quad  HP^{\dagger}+\frac12=0.
 \eea
The second and third equations tell us $P=P^\dagger$ and that $P=-\frac{1}{2H}$. So substituting into the first equation we get 
\be
H^4-mH^2-\frac14=0\quad \Rightarrow H^2=\frac{m+\sqrt{m^2+1}}{2}.
\ee
We see that, while for $m>0$ we still get two gapped vacua, we get two gapped vacua also when $m<0$, $H=\pm\sqrt{\frac{m+\sqrt{m^2+1}}{2}}$ and $P=\mp \sqrt{\frac{2}{m+\sqrt{m^2+1}}}$. Thefore we can see that the linear monopole deformation of the superpotential lifts the $S^1$ vacuum to two points.

\bibliographystyle{ieeetr}

\bibliography{Non_SUSY_dualities}

\end{document}